\documentclass[lettersize,journal]{IEEEtran}
\usepackage{amsmath,amsfonts}
\usepackage{algorithmic}
\usepackage[ruled,noline]{algorithm2e}
\usepackage{array}
\usepackage{tabularx}
\usepackage{makecell}  
\usepackage{textcomp}
\usepackage[caption=true,font=footnotesize,textfont=rm]{subfig}
\usepackage{subfloat}
\usepackage{textcomp}
\usepackage{tikz}

\usepackage{amssymb}
\usepackage{bbding}
\usepackage{pifont}
\usepackage{url}
\usepackage{fancyhdr}
\usepackage{hyperref}
\usepackage{verbatim}
\usepackage{graphicx}
\usepackage{multirow}
\usepackage{cite}
\usepackage{booktabs}
\usepackage{ulem}
\usepackage{soul}
\usepackage{color, xcolor}
\usepackage{float}
\usepackage{makecell}
\usepackage{orcidlink}
\usepackage{tikz}
\usepackage[misc]{ifsym}
\usepackage[switch]{lineno}
\usepackage{cuted} \stripsep -3pt plus 3pt minus 2pt
\usepackage{capt-of}
\begin{document}
\title{DiM-Gestor: Co-Speech Gesture Generation with Adaptive Layer Normalization Mamba-2}

\author{Fan Zhang\hspace{-1.5mm}$^{~\orcidlink{0000-0002-9534-1777}}$,
Siyuan Zhao\hspace{-1.5mm}$^{~\orcidlink{0009-0009-5831-8699}}$,
Naye Ji \hspace{-1.5mm}$^{~\orcidlink{0000-0002-6986-3766}}$,
Zhaohan Wang\hspace{-1.5mm}$^{~\orcidlink{0009-0005-4783-6213}}$,
Jingmei Wu \hspace{-1.5mm}$^{~\orcidlink{0009-0000-9221-8784}}$,
Fuxing Gao\hspace{-1.5mm}$^{~\orcidlink{0009-0008-5586-4734}}$,
Zhenqing Ye\hspace{-1.5mm}$^{~\orcidlink{0009-0003-4341-2734}}$, 
Leyao Yan \hspace{-1.5mm}$^{~\orcidlink{0009-0008-8797-176}}$,
Lanxin Dai \hspace{-1.5mm}$^{~\orcidlink{ 0009-0001-1516-6180}}$,
Weidong Geng \hspace{-1.5mm}$^{~\orcidlink{0000-0002-2709-396X}}$,
Xin Lyu \hspace{-1.5mm}$^{~\orcidlink{0009-0000-1055-6334}}$,
Bozuo Zhao \hspace{-1.5mm}$^{~\orcidlink{0009-0008-2152-0087}}$,
Dingguo Yu\hspace{-1.5mm}$^{~\orcidlink{0000-0002-7674-444X}}$,
Hui Du\hspace{-1.5mm}$^{~\orcidlink{0000-0001-6551-2064}}$, 

Bin Hu $^{(\textrm{\Letter})}$\hspace{-1.5mm}$^{~\orcidlink{0009-0008-0112-5354}}$ %

% ~\IEEEmembership{Staff,~IEEE,}
\thanks{Fan Zhang, Naye Ji, Fuxing Gao, Zhenqing Ye, Leyao Yan, Lanxin Dai, Dingguo Yu, Hui Du, are with the School of Media Engineering, Communication University of Zhejiang, China; (e-mail: fanzhang@cuz.edu.cn; jinaye@cuz.edu.cn; fuxing@cuz.edu.cn; zhenqingye@stu.cuz.edu.cn; leyaoyan@stu.cuz.edu.cn; lanxindai@stu.cuz.edu.cn; yudg@cuz.edu.cn; duhui@cuz.edu.cn; )}
\thanks{Jingmei Wu is with the School of Broadcast Announcing Arts, Communication University of Zhejiang, China; (e-mail: 20190095@cuz.edu.cn;)}
\thanks{Siyuan Zhao, Bin Hu, are with the Faculty of Humanities and Arts, Macau University of Science and Technology, Macau, China (e-mail: 2109853jai30001@student.must.edu.mo, binhu@must.edu.mo)}
\thanks{Zhaohan Wang, Xin Lyu are with the School of Animation and Digital Arts
Communication University of China, Beijing, China (e-mail: 2022201305j6018@cuc.edu.cn; lvxinlx@cuc.edu.cn) }
\thanks{Weidong Geng is with the College of Computer Science and Technology, Zhejiang University, the Research Center for Artificial Intelligence and Fine Arts, Zhejiang Lab, Zhejiang, China (e-mail: gengwd@zju.edu.cn)}
\thanks{Zhao Bozuo is with Changjiang Academy of Art and Design, Shantou University, China (e-mail: bzzhao@stu.edu.cn)}
\thanks{We would like to thank Jiayang Zhu, Weifan Zhong, Huaizhen Chen, and Qiuyi Shen from the College of Media Engineering at the Communication University of Zhejiang for their invaluable technical support in recording the dataset. Additionally, we extend our thanks to Xiaomeng Ma, Yuye Wang, Yanjie Cai,  Xiaoran Chen, Jinyan Xiao, Jialing Ma, Zicheng He, Shuyang Fang, Shuyu Fang, Shixue Sun, Shufan Ma, Sen Xu, Jiabao Zeng, Yue Xu, and Senhua He from the School of Broadcast Arts and School of International Communication \& Education at Communication University of Zhejiang for their contribution as professional TV broadcasters. This work was partially supported by the Pioneer and Leading Goose R\&D Program of Zhejiang (No.2023C01212), the Public Welfare Technology Application Research Project of Zhejiang (No.LGF21F020002, No.LGF22F020008), the National Key Research and Development Program of China (No.2022YFF 0902305).}
\thanks{Code and dataset can be accessed at \href{https://github.com/zf223669/DiMGestures} {https://github.com/zf223669/ DiMGestures.}}

}

\maketitle
% \begin{strip}\centering
% \includegraphics[width=\textwidth]{Figures/chinese_host.png}
% \captionof{figure}{We propose \textit{DiM-Gestor}, an end-to-end AdaLN Mamba-2 and diffusion-based architecture for co-speech gesture generation. In addition, we present the comprehensive Chinese Co-Speech Gestures (CCG) dataset, comprising 15.97 hours of full-body gesture motion performed by professional Chinese TV broadcasters. This dataset encompasses six distinct styles across five scenarios.
% \label{fig:recorded}}

% \end{strip}

\begin{abstract}
 Speech-driven gesture generation using transformer-based generative models represents a rapidly advancing area within virtual human creation. However, existing models face significant challenges due to their quadratic time and space complexities, limiting scalability and efficiency. To address these limitations, we introduce \textit{DiM-Gestor}, an innovative end-to-end generative model leveraging the Mamba-2 architecture. \textit{DiM-Gestor} features a dual-component framework: (1) a fuzzy feature extractor and (2) a speech-to-gesture mapping module, both built on the Mamba-2. The fuzzy feature extractor, integrated with a Chinese Pre-trained Model and Mamba-2, autonomously extracts implicit, continuous speech features. These features are synthesized into a unified latent representation and then processed by the speech-to-gesture mapping module. This module employs an Adaptive Layer Normalization (AdaLN)-enhanced Mamba-2 mechanism to uniformly apply transformations across all sequence tokens. This enables precise modeling of the nuanced interplay between speech features and gesture dynamics. We utilize a diffusion model to train and infer diverse gesture outputs. Extensive subjective and objective evaluations conducted on the newly released Chinese Co-Speech Gestures dataset corroborate the efficacy of our proposed model. Compared with Transformer-based architecture, the assessments reveal that our approach delivers competitive results and significantly reduces memory usage—approximately 2.4 times—and enhances inference speeds by 2 to 4 times. Additionally, we released the CCG dataset, a Chinese Co-Speech Gestures dataset, comprising 15.97 hours (six styles across five scenarios) of 3D full-body skeleton gesture motion performed by professional Chinese TV broadcasters. 
 
\end{abstract}


\section{Introduction}
\begin{figure*}
    \centering
    \includegraphics[width=\linewidth]{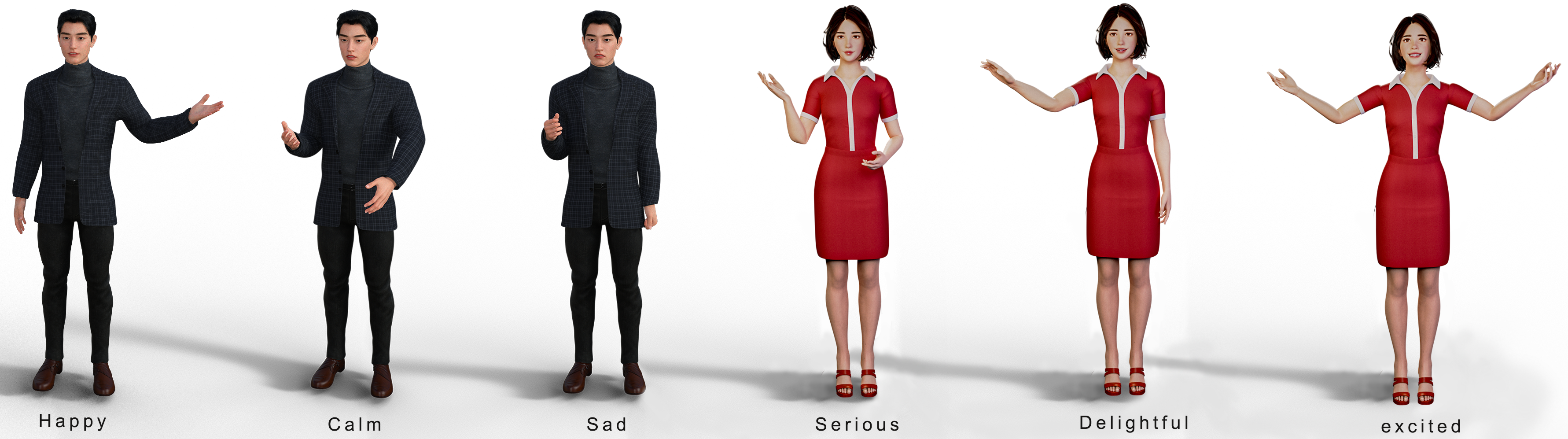}
    \caption{We propose \textit{DiM-Gestor}, an end-to-end AdaLN Mamba-2 and diffusion-based architecture for co-speech gesture generation. In addition, we present the comprehensive Chinese Co-Speech Gestures (CCG) dataset, comprising 15.97 hours of full-body gesture motion performed by professional Chinese TV broadcasters. This dataset encompasses six distinct styles across five scenarios.}
    \label{fig:recorded}
\end{figure*}
\IEEEPARstart{R}{ecent} advancements in 3D virtual human technology have broadened its applications across sectors like animation, human-computer interaction, and digital hosting. A key focus is generating realistic, personalized co-speech gestures, now made feasible through deep learning. Speech-driven gesture generation offers a cost-efficient, automated alternative to traditional motion capture, reducing manual effort while enhancing the realism and adaptability of virtual avatars for diverse professional and recreational uses.

Achieving gesture-speech synchronization with naturalness remains challenging in speech-driven gesture generation. Transformer and Diffusion-based models have improved efficiency and flexibility, leading to innovations like Diffuse Style Gesture\cite{yang_diffusestylegesture_2023}, Diffuse Style Gesture+\cite{yang2023DiffuseStyleGestureaplus}, GestureDiffuClip\cite{ao2023GestureDiffuCLIP}, and LDA\cite{alexanderson2023listen}. Notably, Persona-Gestor\cite{zhang_speech-driven_2024}, using a Diffusion Transformer (DiT)\cite{peebles2022Scalable} architecture, achieves state-of-the-art results by effectively modeling the speech-gesture relationship. However, its transformer-based design imposes high memory usage and slower inference speeds, underscoring the need for more efficient solutions for real-time applications.

The Mamba architecture\cite{gu_mamba_2023}, addresses the quadratic complexity of traditional transformers. Validated across domains like vision\cite{zhu2024vision,ma2024u,li_videomamba_2024}, segmentation\cite{xing2024segmamba} and image tasks\cite{behrouz2024graph,wang2024graph}. The improved Mamba-2\cite{dao_transformers_2024} confirms Mamba's theoretical equivalence to transformers via State Space Duality (SSD) while reducing complexity to linear. This advancement enables faster, resource-efficient processing, making Mamba a compelling alternative for tasks like speech-driven gesture generation, delivering comparable performance to transformers at reduced computational cost.
 
Training datasets in this field, including Trinity\cite{ferstlylva2018Investigating}, ZEGGS\cite{ghorbani2023ZeroEGGS}, BEAT\cite{liu2022BEAT} and Hands 16.2M\cite{lee_gilwoo_talking_2019}, largely focus on English content. Although BEAT offers 12 hours of Chinese speech data, it primarily features spontaneous speech. It limits its suitability for formal contexts like TV broadcasting or structured dialogues.

In this study, we present \textit{DiM-Gestor}, an innovative model leveraging Mamba-2 and diffusion-based architectures to synthesis personalized gestures. The framework utilizes a Mamba-2 fuzzy feature extractor to autonomously capture nuanced fuzzy features from raw speech audio. \textit{DiM-Gestor} integrates an AdaLN Mamba-2 module within its diffusion-based architecture, effectively modeling the intricate dynamics between speech and gestures. Inspired by DiT \cite{peebles2022Scalable} and PG \cite{zhang_speech-driven_2024}, the inclusion of AdaLN significantly enhances the model's ability to accurately capture and reproduce the complex interplay between speech and gestures. Compared with the adaLN transformer, the adaLN Mamba-2 achieves competitive performance while substantially optimizing resource efficiency, reducing memory usage by approximately 2.4 times, and improving inference speeds by a factor of 2 to 4.
 
Further, we released the \textit{CCG} dataset, comprising 15.97 hours of 3D full-body skeleton gesture motion, encompassing six styles across five scenarios performed by professional Chinese TV broadcasters. This dataset provides high-quality, structured data, facilitating advanced research in Chinese speech-driven gesture generation, particularly for applications requiring formal and contextually appropriate non-verbal communication, as shown in Figure \ref{fig:recorded}. 

For clarity, our contributions are summarized as follows:
\begin{itemize}
    \item{\textbf{We introduce \textit{DiM-Gestor}, an innovative end-to-end generative model leveraging the Mamba-2 architecture:}} Our approach achieves competitive performance while significantly optimizing resource efficiency, reducing memory usage by approximately 2.4 times and enhancing inference speeds by a factor of 2 to 4.
    \item{\textbf{We released the CCG dataset, a Chinese Co-Speech Gestures dataset:}} This comprehensive dataset, captured using inertial motion capture technology, comprises 15.97 hours of 3D full-body skeleton gesture motion. It includes six distinct styles across five scenarios, performed by professional Chinese TV broadcasters, offering high-quality, structured data for advancing research in Chinese speech-driven gesture synthesis.

    \item{\textbf{Extensive subjective and objective evaluations:}} These evaluations demonstrate that our model surpasses current state-of-the-art methods, highlighting its exceptional capability to generate credible, speech-appropriate, and personalized gestures while achieving reduced memory consumption and faster inference times.

\end{itemize}

\section{RELATED WORK}\label{sec:RELATED WORK}
This section briefly overviews transformer- and diffusion-based generative models for speech-driven gesture generation.
\subsection{Transformer- and diffusion-based generative models}
DiffMotion \cite{zhang2023diffmotion} represents a pioneering application of diffusion models in gesture synthesis, incorporating an LSTM to enhance gesture diversity. Cross-modal Quantization (CMQ)\cite{wang2024cross}, jointly learns and encodes the quantized codes for representations of the speech and gesture together. However, these model supports only the upper body. Alexanderson et al. \cite{alexanderson2023listen} refined DiffWave by replacing dilated convolutions, thereby unlocking the potential of transformer architectures for gesture generation. GestureDiffuCLIP (GDC)\cite{ao2023GestureDiffuCLIP} employs transformers and AdaIN layers to integrate style guidance directly into the diffusion process. Similarly, DiffuseStyleGesture (DSG) \cite{yang_diffusestylegesture_2023} and its extension DSG+ \cite{yang2023DiffuseStyleGestureaplus} utilize cross-local attention and layer normalization within transformer models. While these methods have demonstrated significant progress, they often struggle to balance gesture and speech synchronization. This leads to gestures that can appear either overly subtle or excessively synchronized with speech.

Persona-Gestor (PG) \cite{zhang_speech-driven_2024} addresses some of these challenges by introducing a fuzzy feature extractor. This approach uses 1D convolution to capture global features from raw speech audio, paired with an Adaptive Layer Normalization (AdaLN) transformer \cite{peebles2022Scalable} to model the nuanced correlation between speech features and gesture sequences. Although Persona-Gestor achieves high-quality motion outputs, it is hindered by substantial memory requirements and slower inference speeds associated with convolutional and transformer-based architectures.

\subsection{Co-speech gesture training datasets}
The datasets commonly employed for training co-speech gesture models include Trinity \cite{ferstlylva2018Investigating,ferstlylvaAdversarialGestureGeneration2020}, ZEGGS \cite{ghorbani2023ZeroEGGS}, BEAT \cite{liu2022BEAT}, and Hands 16.2M \cite{lee_gilwoo_talking_2019}. These datasets predominantly feature native English speakers engaged in spontaneous conversational speech. Although the BEAT dataset includes 12 hours of Chinese content, this subset is characterized by unstructured speech patterns, making it less suitable for applications requiring formal contexts, such as event broadcasting or structured dialogues. While Yoon et al. \cite{yoon_youngwoo_robots_2019} present a dataset collected and extracted the skeleton motions from TED talks videos, this dataset's quality is unsuitable for high-quality gesture synthesis tasks.

To address these limitations, we adopt the Mamba-2 architecture \cite{dao_transformers_2024}, further adapting it with an AdaLN-based implementation. The Mamba-2 framework significantly reduces memory usage and improves inference speed, offering a more efficient solution for gesture synthesis in virtual human interactions. In addition, we introduce the Chinese Co-speech Gestures (CCG) dataset, comprising 15.97 hours of 3D full-body skeleton motion across six gesture styles and five scenarios. The dataset features high-quality performances by professional Chinese TV broadcasters. This dataset enables advanced research in speech-driven gesture generation, particularly in formal settings such as event hosting and professional presentations.

\section{Problem Formulation}
We conceptualize co-speech gesture generation as a sequence-to-sequence translation task, where the goal is to map a sequence of speech audio features, \(X = [x_t]_{t=1}^{t=T} \in \mathbb{R}^{T}\), to a corresponding sequence of full-body gesture features, \(Y^0 = [y^0_t]_{t=1}^{t=T} \in \mathbb{R}^{T \times (D+6)}\). Each gesture frame \(y^0_t \in \mathbb{R}^{(D+6)}\) comprises 3D joint angles, as well as root positional and rotational velocities, with \(D\) representing the number of joint channels and \(T\) denoting the sequence length. This formulation encapsulates the temporal and spatial complexity of mapping audio-driven dynamics to expressive human-like gestures.

We define the probability density function (PDF), \(p_\theta(\cdot)\), to approximate the true gesture data distribution \(p(\cdot)\), enabling efficient sampling of gestures. The objective is to generate a non-autoregressive full-body gesture sequence (\(Y^0\)) from its conditional probability distribution, given the audio sequence (\(X\)) as a covariate:

\begin{equation}  
Y^0 \sim p_\theta(Y^0 \mid X) \approx p(Y^0 \mid X)
\end{equation}

This formulation leverages a denoising diffusion probabilistic model trained to approximate the conditional distribution of gestures aligned with speech. This approach provides a robust framework for learning and synthesizing co-speech gestures with high fidelity and temporal coherence by modeling the intricate relationship between audio inputs and gesture outputs.

\section{System Overview}\label{sec:PROPOSED APPROACH}
\begin{figure*}[!ht]
\centering
\includegraphics[width=\textwidth]{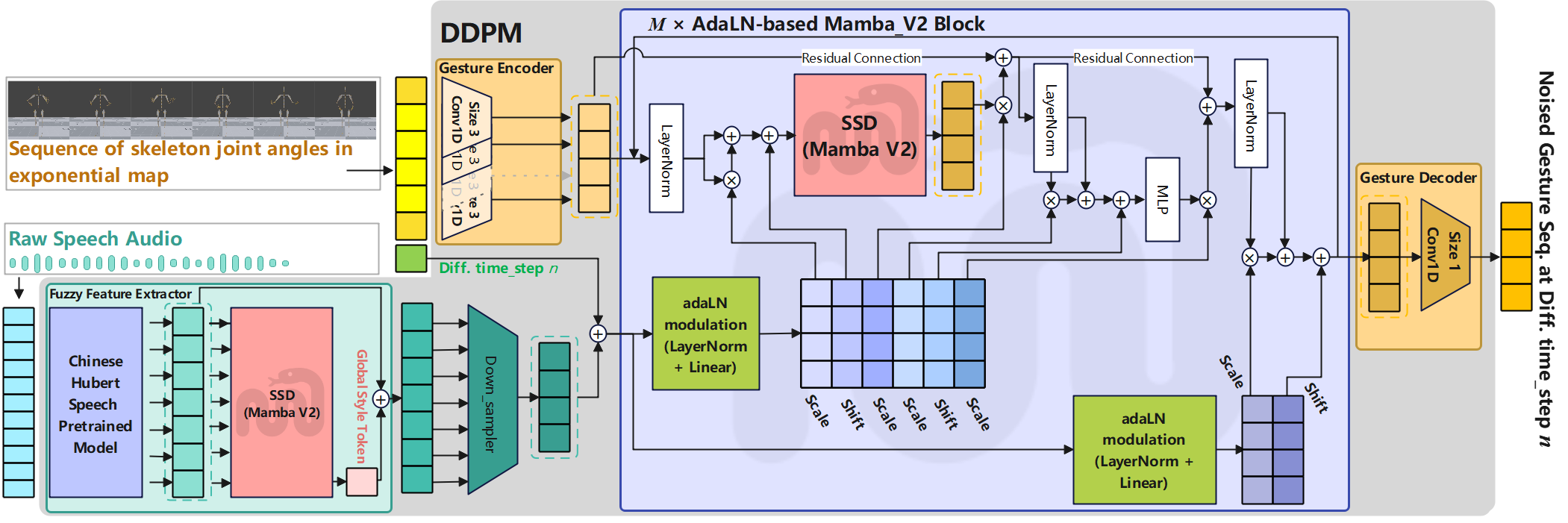}
% \label{fig:architecture}
\caption{The architecture of DiM-Gestor incorporates a Mamba-2 fuzzy feature extractor and an Adaptive Layer Normalization (AdaLN) Mamba-2 diffusion architecture. The fuzzy feature extractor features a dual-component system engineered to capture the nuanced style and detailed audio features into unified latent features. These unified latent features are subsequently channeled into the AdaLN Mamba-2. This module is pivotal in modeling the intricate relationship between the incoming audio features and the corresponding gestures. It facilitates the estimation of diffusion noise within the diffusion model, ensuring the generation of diverse gestures. The overall schematic includes three main components: (a) Mamba-2 Fuzzy Feature Extractor, (b) Stack of AdaLN Mamba-2 Blocks, (c) Gestures Encoder and Decoder, and (d) Denoising Diffusion Probabilistic Model (DDPM).}
\label{fig:architecture}
\end{figure*}

DiM-Gestor, an end-to-end Mamba- and diffusion-based architecture, processes raw speech audio as sole input to synthesize personalized gestures. This model balances naturalness with lower resource consumption, as shown in Figure \ref{fig:architecture}.
% %%%%%%%%%%%%%%%%% poses, gestures, identity
%%%%%%%%%%%%%%%%% Figure 1 %%%%%%%%%%%%%%%%%%%%%%%%%%%%
%%%%%%%%%%%%%%%%% Figure 1 %%%%%%%%%%%%%%%%%%%%%%%%%%%%

\subsection{Model Architecture}
The architecture of \textit{DiM-Gestor}, depicted in Figure \ref{fig:architecture}, integrates four key components to efficiently generate personalized gestures directly from speech audio: \textbf{(a) Mamba-2 Fuzzy Feature Extractor}: This module employs a fuzzy inference strategy utlizing Mamba-2 to autonomously capturing nuanced stylistic and contextual elements from raw speech audio. \textbf{(b) Stack of AdaLN Mamba-2 Blocks}: These blocks introduce AdaLN Mamba-2 architecture that applies uniform transformations across all tokens, enabling the model to effectively capture the intricate interplay between speech and gestures while enhancing computational efficiency. \textbf{(c) Gestures Encoder and Decoder}: These modules encode gesture sequences into latent representations and decode them back into full-body motion outputs, ensuring accurate reconstruction of gesture dynamics. \textbf{(d) Denoising Diffusion Probabilistic Model (DDPM)}: As the backbone for probabilistic generation, this module leverages diffusion processes to synthesize diverse and realistic gesture sequences aligned with the given speech context.

By combining these components into a unified framework, the \textit{DiM-Gestor} architecture captures the complexity of human gestures in relation to speech while significantly reducing memory consumption and improving inference speed. This design ensures high-quality, personalized, and contextually appropriate gesture generation. 

\subsubsection{Mamba-2 Fuzzy Feature Extractor}
This module employs a fuzzy inference strategy, which does not rely on explicit and manual classification inputs. Instead, it provides implicit, continuous, and fuzzy feature information, automatically learning and inferring the global style and specific details directly from raw speech audio. Illustrated in Figure \ref{fig:architecture}, this module is a dual-component extractor comprising both global and local extractors. The local extractor utilizes the Chinese Hubert Speech Pretrained Model \cite{TencentGameMate} to process the audio sequence into discrete tokens. This pre-trained model, for its proficiency in capturing the complex attributes of speech audio, allows it to effectively represent universal Chinese speech audio latent features, denoted as $Z_x$.

We implement a Mamba-2 \cite{dao_transformers_2024} global style extractor framework. In the Mamba architecture, the module scans the entire sequence of $Z_x$ to capture the style feature. The last output token,  $z_s$, is considered crucial as it encompasses the global style feature contained within the speech audio. This feature is then broadcast to align with the local fuzzy features, ensuring that the global context influences the local gesture synthesis. This process allows the model to maintain a holistic understanding of the style and emotional context throughout the gesture generation process. This unified latent representation is then channeled to the downsampling module for further refinement.

The downsampling module, crucial for aligning each latent representation with its corresponding sequence of encoded gestures, is seamlessly integrated into the condition extractor. We implement a Conv1D layer with a kernel size of 201 within our architecture\cite{zhang_speech-driven_2024}. The size of the kernel in the gesture synthesis model suggests that the preceding gesture is influenced not only by the current semantics but also by the prior and subsequent semantic contexts \cite{mcneill1992Hand}. This kernel size thus plays a crucial role in capturing the temporal dynamics across sequences, allowing for a more coherent and contextually integrated gesture generation that aligns with the natural flow of speech and its semantic shifts. The output of this module, $C=[c_t]_{t=1}^{t=T}$, serves as a unified latent representation that encapsulates both encoded acoustic features and the diffusion time step $n$, ensuring a coherent and accurate gesture generation process. 

% \begin{figure}[htbp]
% \includegraphics[width=0.45\textwidth]{Figures/Fuzzy_Feature_Extractor.png}
% \begin{center}
% \caption{An overview of the Mamba-based style fuzzy inference condition extractor. }
% \label{fig:ConditionEncoder} 
% \end{center}
% \end{figure}

\subsubsection{AdaLN Mamba-2}
The AdaLN's fundamental purpose is to incorporate a conditional mechanism \cite{peebles2022Scalable, zhang_speech-driven_2024} that uniformly applies a specific function across all tokens. It offers a more sophisticated and nuanced approach to modeling, enabling the system to capture and articulate the complex dynamics between various input conditions and their corresponding outputs. Consequently, this improves the model's predictive accuracy and ability to generate outputs more aligned with the given conditions. 

Diffusion Transformers (DiTs) exemplify a sophisticated advancement in diffusion model architectures, incorporating an AdaLN-infused transformer framework primarily for text-to-image synthesis. The utility of DiTs has recently expanded to include text-conditional video generation, illustrating their versatility. Furthermore, DiTs have shown potential in co-speech gesture generation \cite{zhang_speech-driven_2024}, marking a significant step in applying these models to sequence-based tasks. However, the inherent quadratic space complexity associated with Transformers results in substantial memory consumption and slower inference speeds. 

The AdaLN architecture involves regressing the dimension-wise scale and shift parameters (\(\gamma\in(T,D)\) and \(\beta\in(T,D)\)), derived from the fuzzy feature extractor output \(C\), rather than directly learning these parameters, as depicted in Figure \ref{fig:architecture} and algorithm \ref{alg:adaln}. In each AdaLN Mamba-2 stack, a latent feature \(z^n_{1:T,m}\) is generated, combining condition information and gesture using AdaLN and the Mamba-2 architecture. The index \(m\) ranges from 1 to \(M\), where \(M\) represents the total number of AdaLN Mamba-2 stacks. Furthermore, as illustrated in Figure \ref{fig:architecture}, the final layer utilizes the same fuzzy features, supplemented by a scale and shift operation to fine-tune the gesture synthesis.

\begin{algorithm}[htbp]
\SetKwFor{For}{for}{do}{end\enspace for}
\SetKwIF{If}{ElseIf}{Else}{if}{then}{else if}{else}{end\enspace if}
\SetKw{Return}{Return:}
\caption{Adaptive Layer Normalization Mamba-2}
\KwIn{ Encoded Gestures $G_{1:T}$ and Conditional features $C_{1:T}$} 
\For {$m = 0$ \emph{\KwTo} $M-1$}{
  $\gamma_1, \beta_1, \alpha_1, \gamma_2, \beta_2, \alpha_2 = MLP(C_{1:T}).Chunk(6) $ \\
  $X_{1:T} = X_{1:T} + \alpha_1 × Mamba2(LN(G_{1:T}) × (1+\gamma_1) + \beta_1)$ \\
  $X^{m+1}_{1:T} = X_{1:T} + \alpha_2 × MLP(LN(X^{m}_{1:T}) × (1+\gamma_2) + \beta_2))$
}
  $\gamma_3, \beta_3 = MLP(C_{1:T}).Chunk(2)$ \\
  $Z_{1:T} = LN(X_{1:T}) × (1+\gamma_3) + \beta_3)$ \\
\Return{$Z_{1:T}$}
\label{alg:adaln}
\end{algorithm}

% %%%%%%add mamba-2 description
% In contrast to traditional Diffusion Transformers, our approach integrates the Mamba-2 architecture \cite{dao_transformers_2024} as a replacement for the conventional transformer module. This strategic adaptation leverages the minimal memory footprint and enhanced processing efficiency of Mamba-2, significantly accelerating inference speeds without sacrificing output quality. This novel substitution is pivotal in addressing the challenges associated with speech-driven gesture synthesis, improving efficient and high-quality performance.

Mamba-2\cite{dao_transformers_2024}, an evolution within the Structured State Space Duality (SSD) framework, refines Mamba's selective state-space model (SSM), markedly enhancing computational efficiency. This architecture is engineered to supplant the traditional attention mechanism in Transformers, specifically utilizing structured semiseparable matrices to optimize facets such as training speed and memory consumption. Its robust performance on modern hardware, versatility across varying sequence lengths, and the implementation of tensor parallelism firmly establish Mamba-2 as a potent alternative to traditional attention-based models, offering substantial efficiency gains and reduced operational costs.

The dual form of Structured State Space Duality (SSD) is characterized by a quadratic computation closely related to the attention mechanism. It can be defined as follows:

\begin{equation}  
(L \circ (QK^T)) \cdot V \quad L_{ij} = 
\begin{cases} 
a_i \times \cdots \times a_{j+1} & \text{if } i \geq j \\
0 & \text{if } i < j 
\end{cases}
\end{equation}

where \(a_i\) are input-dependent scalars bounded within the interval \([0,1]\). These scalars represent the normalized attention weights in the structured state-space model, modulating the influence of each input token on the computed output. For indexing of $a$, \(i:j\) refers to the range \((i, i+1, \dots, j-1)\) when \(i < j\) and \((i, i-1, \dots, j+1)\) when \(i > j\).

\begin{figure}[!htbp]
    \centering
    \includegraphics[width=\linewidth]{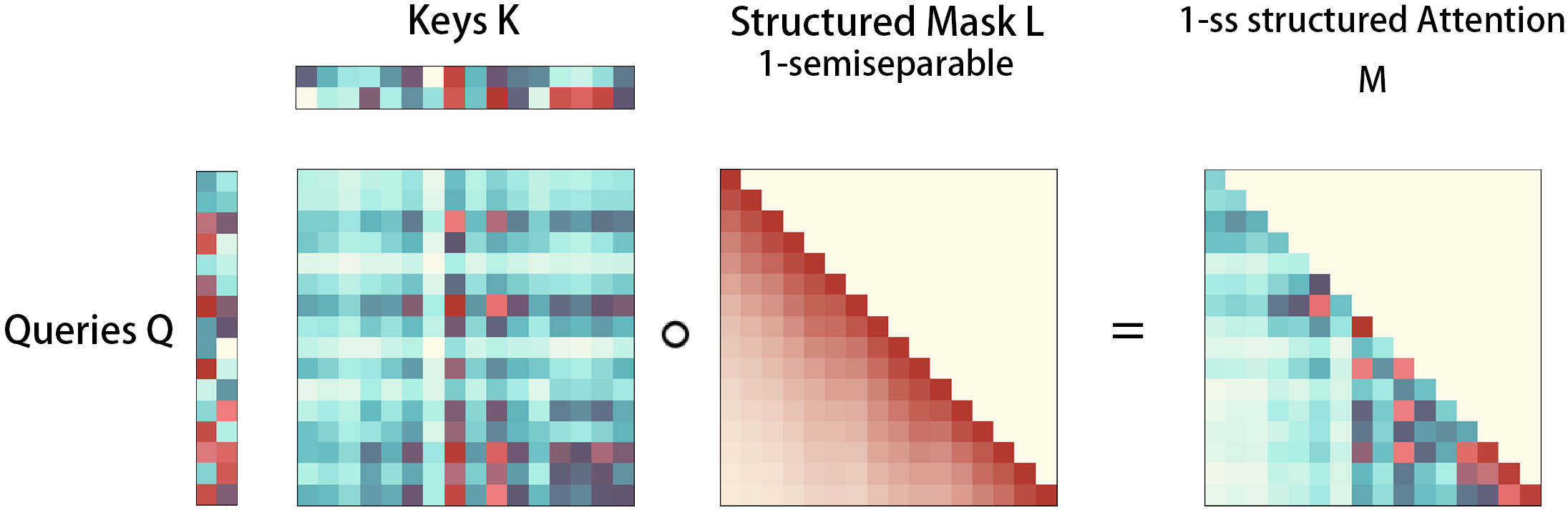}
    \caption{The Structured Masked Attention.}
    \label{fig: attention}
\end{figure}

Let Structured Masked Attention (SMA) $M = (L \circ (QK^T))$, we chose 1-semi separable for structured Mask $L$ for constructing the 1-SS Structured Attention, as shown in Figure \ref{fig: attention}. So the Mamba-2 output is $Ma_n = M \cdot V $, Where the subscript $n$ represents the n-th blocks of the AdaLN Mamba-2.

Compared to standard (self-)attention mechanisms, the Structured State Space (SSD) model with 1-semi separable Structured Masked Attention (SMA) introduces significant optimizations. SSD eliminates the softmax normalization step, effectively reducing the requisite state size of the model from a linear to a constant scale, thereby enhancing computational efficiency from quadratic to linear. Additionally, SSD incorporates a distinct elementwise mask matrix that is applied multiplicatively, further refining the model's efficiency and operational dynamics. This alteration not only simplifies the computational process but also improves the scalability and speed of the model, making it more adept at handling larger datasets and complex tasks without compromising performance. In contrast to the Persona-Gestor \cite{zhang_speech-driven_2024}, which utilizes a causal mask, the primary SSD model employs a 1-Semiseparable Structured Masked Attention (1-SS mask), offering a more structured approach to attention that enhances both computational efficiency and performance in sequence modeling tasks. 

\subsubsection{Gesture Encoder and Decoder}
The architecture of the gesture encoder and decoder is meticulously designed to process gesture sequences, as illustrated in Fig. \ref{fig:architecture}. 

Gesture Encoder employs a Convolution1D layer with a kernel size of 3 to encode the initial sequence of gestures, denoted as \(X\), into a hidden state \(H_x = [h_t]_{t=1}^{t=T}\). Experimental results reveal that using a kernel size of 1 often leads to animation jitter, likely due to insufficient spatial-temporal feature capture. Conversely, the kernel size 3 effectively mitigates this issue by capturing spatial-temporal relationships across adjacent frames, ensuring smoother gesture dynamics.

The decoder transforms the high-dimensional output from the AdaLN Mamba-2 layer to the original dimension, corresponding to the skeletal joint channels. This step involves generating the predicted noise (\(\epsilon_\theta\)), a critical component for gesture reconstruction. Utilizing a 1D kernel of size 1 in the decoder enables the model to extract relevant features and correlations between adjacent joint channels, thereby improving the overall quality and coherence of the generated gestures.

 % instead of employing a fully connected layer.

\subsection{Training and Inferencing with DDPM} \label{sec:DDPM}
The diffusion process, leveraging the Denoising Diffusion Probabilistic Model (DDPM) \cite{zhang2023diffmotion,zhang_speech-driven_2024,ho_denoising_2020,sohl-dickstein_deep_2015}, enabling the reconstruction of the conditional probability distribution between gestures and fuzzy features. 

The model functions through two primary processes: the diffusion process and the generation process. During training, the diffusion process progressively transforms the original gesture data (\(X^0\)) into white noise (\(X^N\)) by optimizing a variational bound on the data likelihood. This gradual addition of noise is meticulously controlled to preserve the underlying data structure, facilitating effective learning of the conditional probability distribution.

During inference, the generation process seeks to reverse the transformation performed during training. It reconstructs the original gesture data from noise by reversing the noising process through a Markov chain, employing Langevin sampling \cite{paul_sur_1908}. This approach facilitates the accurate and effective recovery of gesture data from its perturbed state. 

The Markov chains utilized in the diffusion and generation processes ensure a coherent and systematic transition between stages, thereby preserving the integrity and quality of the synthesized gestures. The specific Markov chains employed in the diffusion and generation processes are as follows:

\begin{equation}
\begin{aligned}
&p\left(Y^n|Y^0\right) = \mathcal{N}\left(Y^n; \sqrt{\overline{\alpha}^n} Y^0, \left(1-\overline{\alpha}^n\right)I\right)   \quad and\\ 
&p_\theta\left(Y^{n-1}|Y^n, Y^0\right) = \mathcal{N}\left(Y^{n-1}; \tilde{\mu}^n\left(Y^n, Y^0\right), \tilde{\beta}^n I\right),
\end{aligned}
\label{eq:cumulativeProduct}
\end{equation}
where \(\alpha^n := 1 - \beta^n\) and \(\overline{\alpha}^n := \prod_{i=1}^n \alpha^i\). As shown by\cite{ho_denoising_2020}, \(\beta^n\) is a increasing variance schedule \(\beta^1,...,\beta^N\) with \(\beta^n \in (0,1)\), and \(\tilde{\beta}^n := \frac{1-\overline{\alpha}^{n-1}}{1-\overline{\alpha}^n}\beta^n\).

The training objective is to optimize the model parameters \(\theta\) by minimizing the Negative Log-Likelihood (NLL). This optimization is implemented through a Mean Squared Error (MSE) loss, which measures the deviation between the true noise, denoted as \(\epsilon \sim \mathcal{N}(0, I)\), and the predicted noise \(\epsilon_\theta\). The objective function can be expressed as:

\begin{equation}
\mathcal{L}_{\text{MSE}} = \mathbb{E}_{x^0, \epsilon, n}\left[\|\epsilon - \epsilon_\theta(x^n, n)\|^2_2\right],
\end{equation}

where \(x^0\) represents the original data, \(x^n\) is the noised version of the data at step \(n\), and \(n\) indicates the diffusion time step. This formulation ensures the model learns to accurately predict noise across all diffusion steps, thereby enabling the effective reconstruction of the original data during inference.

\begin{equation}
\label{eq:objective2}
\mathbb{E}_{Y^0_{1:T}, \epsilon, n}[||\epsilon - \epsilon_\theta\left(\sqrt{\overline{\alpha}^n X^0}+\sqrt{1-\overline{\alpha}^n}\epsilon , X,n\right)||^2],
\end{equation}  Here \(\epsilon_\theta\) is a neural network (see figure \ref{fig:architecture}), which uses input \(Y^0\) , \(X\) and \(n\) that to predict the \(\epsilon\), and contains the similar architecture employed in \cite{rasul_autoregressive_2021}.

After completing the training phase, we utilize variational inference to generate a full sequence of new gestures that align with the original data distribution, formulated as \(Y^0 \sim p_\theta(Y^0 \mid X)\). In this generation phase, the entire sequence \(Y^0\) is sampled from the learned conditional probability distribution, ensuring that the synthesized gestures accurately reflect the dynamics and nuances of the input speech features \(X\).

The term \(\sigma_\theta\) denotes the standard deviation of the conditional distribution \(p_\theta(Y^{n-1} \mid Y^n)\), playing a pivotal role in capturing the variability and intricacies of transitions across diffusion stages. In our model, we define \(\sigma_\theta := \tilde{\beta}^n\), where \(\tilde{\beta}^n\) represents a predetermined scaling factor. This factor adjusts the noise level at each diffusion step, enabling a controlled balance between smoothing and preserving fine details during the generation process.

During inference, we send the entire sequence of raw audio to the condition extractor component. The output of this component is then fed to the diffusion model to generate the whole sequence of accompanying gestures (\(Y^0\)).

\section{EXPERIMENTS}\label{sec:EXPERIMENTS}
Our experiments focused on producing full 3D body gestures, including finger motions and locomotion, trained on our released Chinese Co-speech Gestures (CCG) dataset.

\subsection{Dataset Recording and Data Processing}

\subsubsection{Datasets}
\begin{figure}[htbp]
\includegraphics[width=0.5\textwidth]{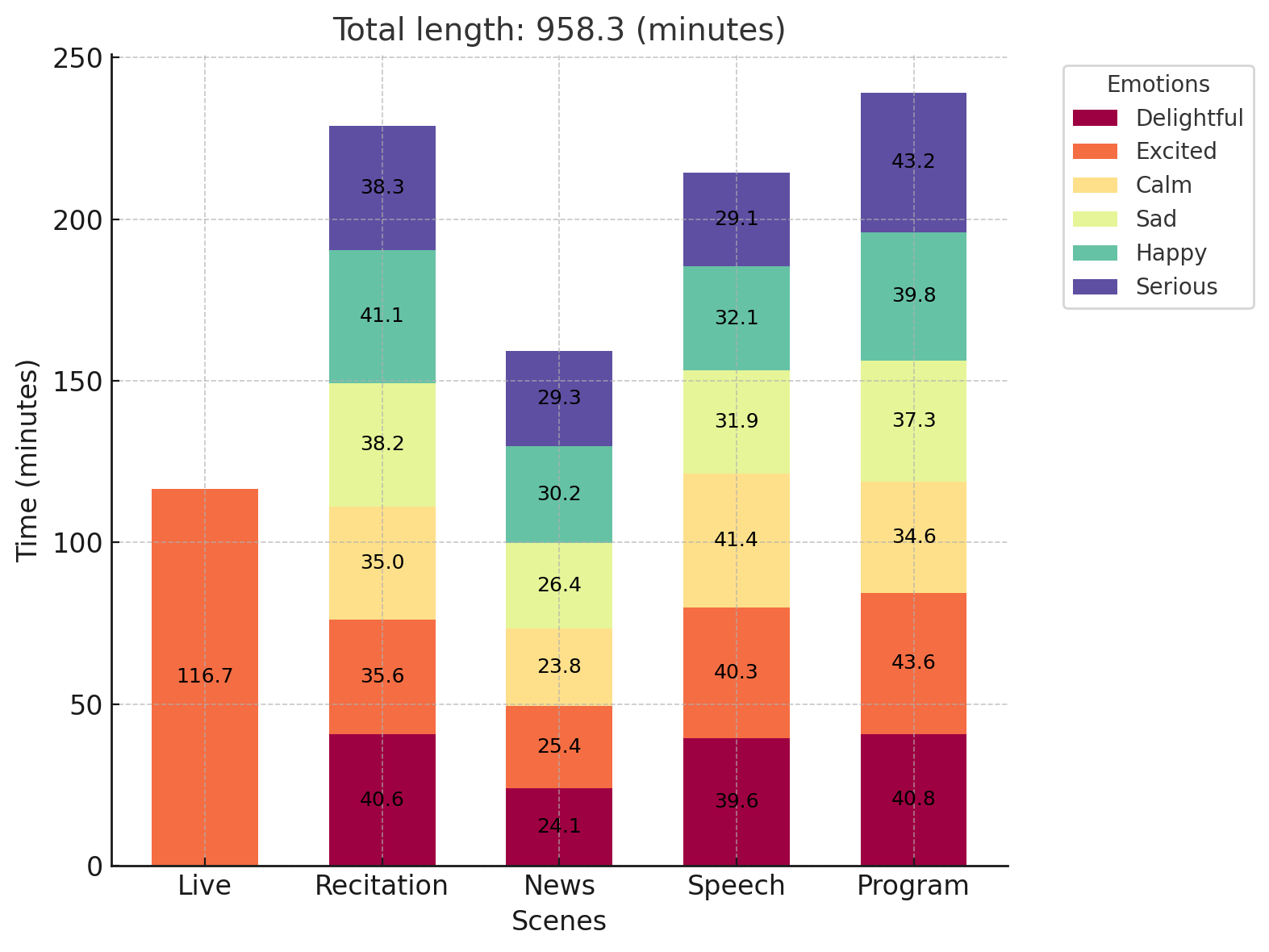}
\begin{center}
\caption{An overview of the recorded Chinese TV broadcasters dataset. }
\label{fig:dataset} 
\end{center}
\end{figure}

We have constructed \textit{Chinese Co-Speech Gestures, CCG} dataset, a high-quality, synchronized motion capture and speech audio dataset comprising 391 monologue sequences. The dataset features performances by 12 female and 5 male actors, all conducted in Chinese and covering 6 distinct emotion styles with 5 different senses. These styles were carefully selected to comprehensively represent various postures, hand gestures, and head movements. The total length of the dataset amounts to 958.3 minutes. Table \ref{tab:dataset_details} and Figure \ref{fig:dataset} illustrate the time distribution of the different motion styles across the various scenes within the dataset.

The style labels for each sequence were assigned according to predefined actor directives. However, it is essential to note that these labels may not always align with the subjective interpretations of independent external annotators regarding the observed movement styles. The motion capture data was recorded using the Noitom PNS system\footnote{https://www.noitom.com.cn/}, which employs inertial motion capture technology.

The full-body motion capture was conducted at a frame rate of 100 frames per second (fps), with the motion data encoded using a skeletal model comprising j=59 joints. Concurrently, the audio data was recorded at a sampling rate of 44,100 Hz.

\begin{table}[h]
\centering
\renewcommand{\arraystretch}{1.2}
\resizebox{0.7\linewidth}{!}{
\begin{tabular}{cc|cc}
\hline
\textbf{Style} & \textbf{\begin{tabular}[c]{@{}c@{}}Length\\ (min)\end{tabular}} & \textbf{Style} & \textbf{\begin{tabular}[c]{@{}c@{}}Length\\ (min)\end{tabular}} \\ \hline
Calm           & 134.72                                                          & Sad            & 133.79                                                          \\
Delightful     & 145.06                                                          & Serious        & 139.92                                                          \\
Excited        & 261.65                                                          & & \\
Happy          & 143.16                                                          &                \textbf{Total}&                                                                 \textbf{958.3}\\ \hline
\end{tabular}
}
\caption{Details of the recorded motion and audio dataset in minutes.}
\label{tab:dataset_details}
\end{table}

\subsubsection{Speech Audio Data Process}
Due to the Chinese Hubert Speech Pretrained Model being pre-trained on speech audio sampled at 16 kHz, we uniformly resampled all audio down from 44.1 kHz to match this frequency, ensuring compatibility and optimal performance.

\subsubsection{Gesture Data Process}
We concentrate exclusively on full-body gestures, employing the data processing techniques detailed by Alexanderson et al. \cite{alexanderson_simon_style-controllable_2020}. This includes capturing translational and rotational velocities to accurately delineate the root's trajectory and orientation. The datasets are uniformly downsampled to a frame rate of 20 fps. To ensure precise and continuous representation of joint angles, we utilize the exponential map technique \cite{grassiaf.sebastian1998Practical}. All data are segmented into 20-second clips for training and validation purposes.

\subsection{Model Settings}
Our experiments utilized Mamba-2 architecture for a global fuzzy feature extractor and six AdaLN Mamba-2 blocks, each Mamba-2 configured with a 256 SSM state expansion factor, a local convolution width of 4, and a block expansion factor of 2. This encoding process transforms each frame of the gesture sequence into hidden states \(h \in \mathbb{R}^{1280}\). We employ the Chinese Hubert Speech Pretrained Model (chinese-wav2vec2-base)\footnote{https://github.com/TencentGameMate/chinese\_speech\_pretrain} for audio processing.

The diffusion model uses a quaternary variance schedule, starting from \(\beta_1 = 1 \times 10^{-4}\) to \(\beta_N = 8 \times 10^{-2}\) with a linear beat schedule, and a total of \(N = 1000\) diffusion steps. The training batch size is set to 32.

Our model was tested on Intel i9 CPU with a Nvidia Geforce 4090 GPU, in contrast to the A100 GPU used by Persona-Gestor \cite{zhang_speech-driven_2024}. The training time was approximately 6 hours.

%%%%%%%%%%%%%%%%%%%%%%%%%%%%%%%%%%%%%%%%%%%%%%%%%%%%%%
\subsection{Visualization Results}
Our system excels in generating personalized gestures that are contextually aligned with speech by leveraging the Mamba-2 fuzzy inference extractor and adaLN Mamba-2.

\begin{figure}[!htbp]
    \centering
    \includegraphics[width=0.75\linewidth]{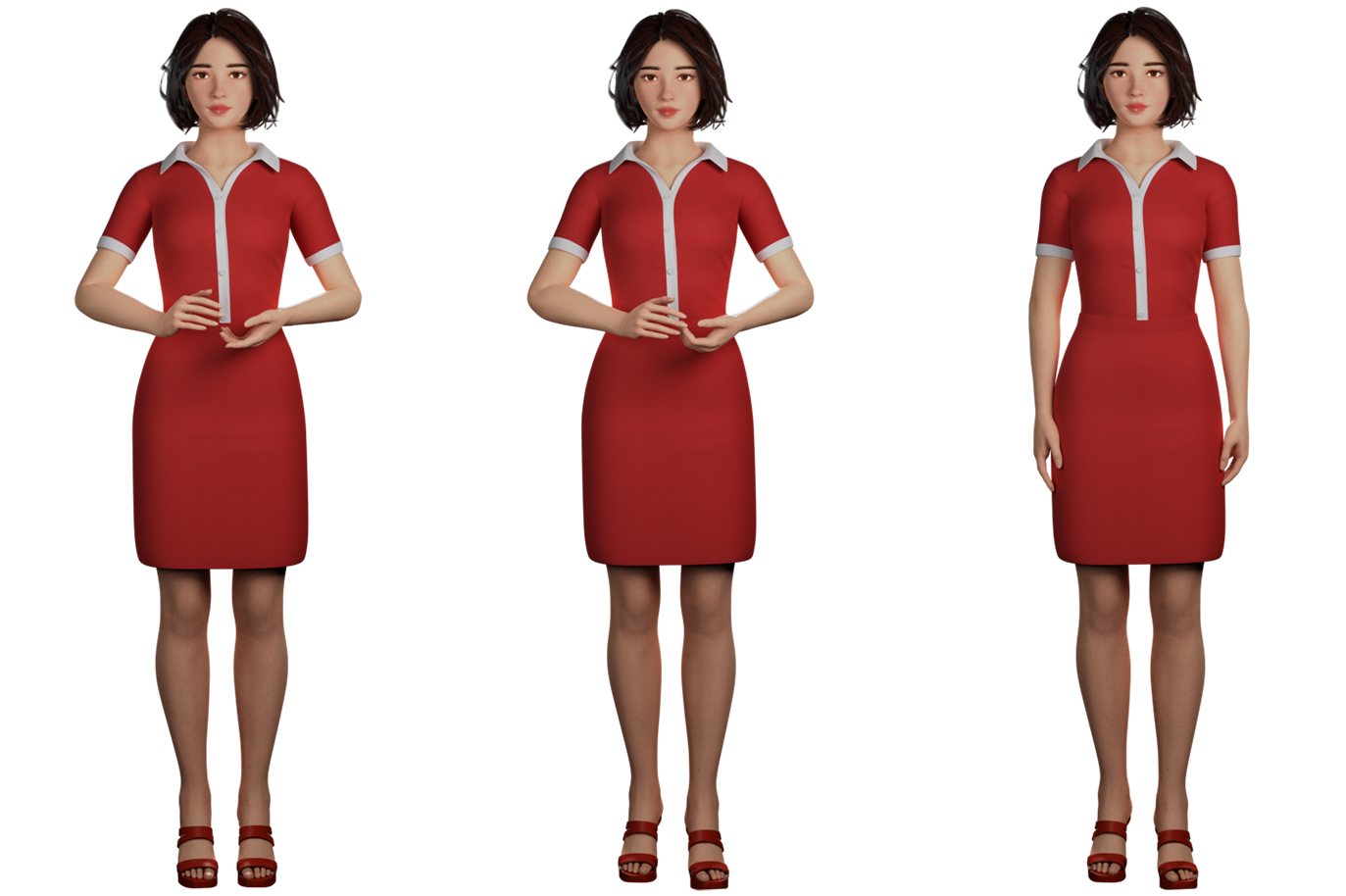}
    \caption{The female broadcaster is presenting neutral programs live on air (Left: GT; Center: DiM-Gestor; Right: PG-12blocks).}
    \label{fig:f_program_calm}
\end{figure}

\begin{figure}[!htbp]
    \centering
    \includegraphics[width=0.75\linewidth]{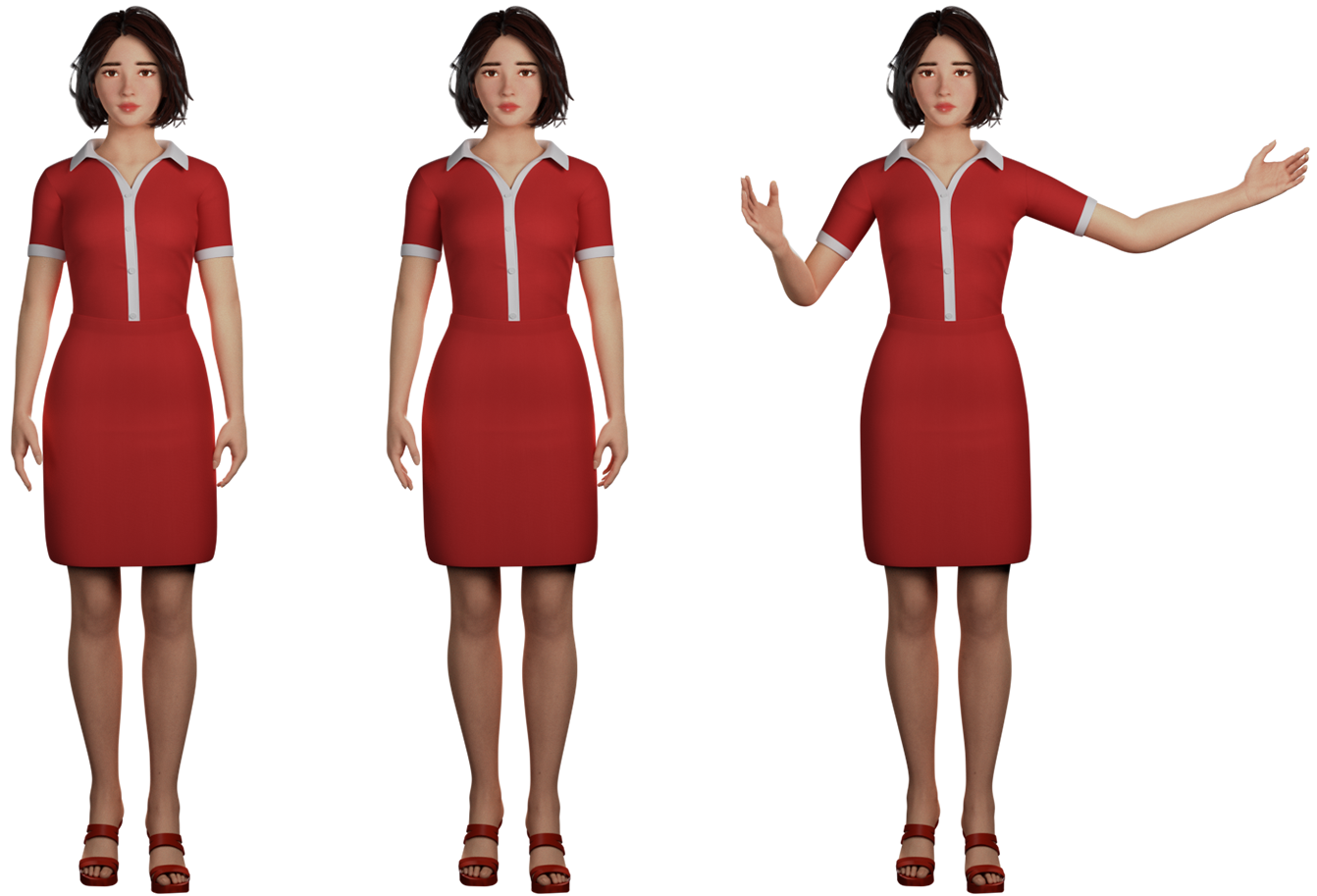}
    \caption{The female broadcaster delivering a recitation on the subject of sad (Left: GT; Center: DiM-Gestor; Right: PG-12blocks).}
    \label{fig:f_recitation_sad}
\end{figure}

\begin{figure}[!htbp]
    \centering
    \includegraphics[width=0.75\linewidth]{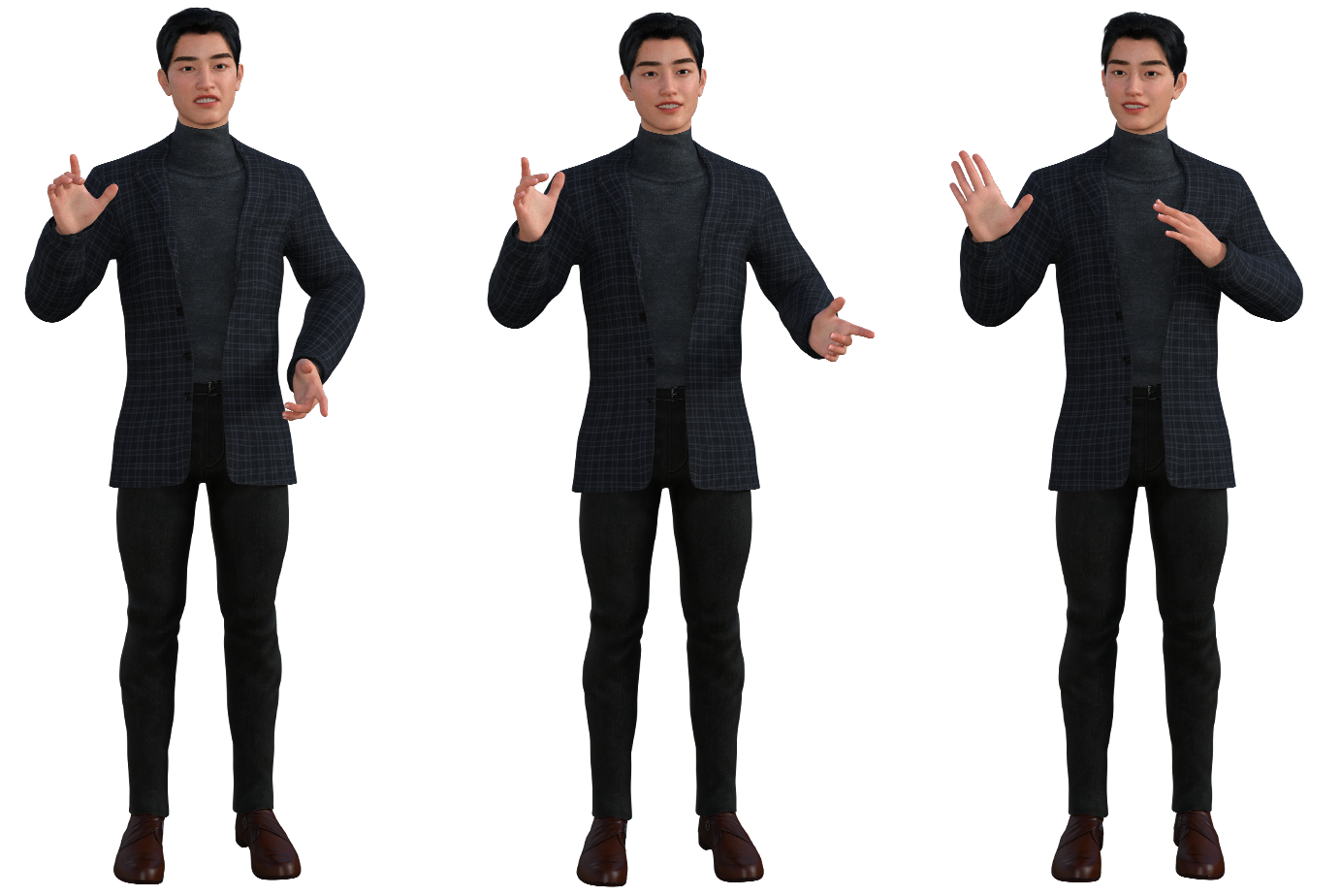}
    \caption{The male broadcaster energetically delivered the exciting news (Left: GT; Center: DiM-Gestor; Right: PG-12blocks). }
    \label{fig:m_news_excited}
\end{figure}

\begin{figure}[!htbp]
    \centering
    \includegraphics[width=0.75\linewidth]{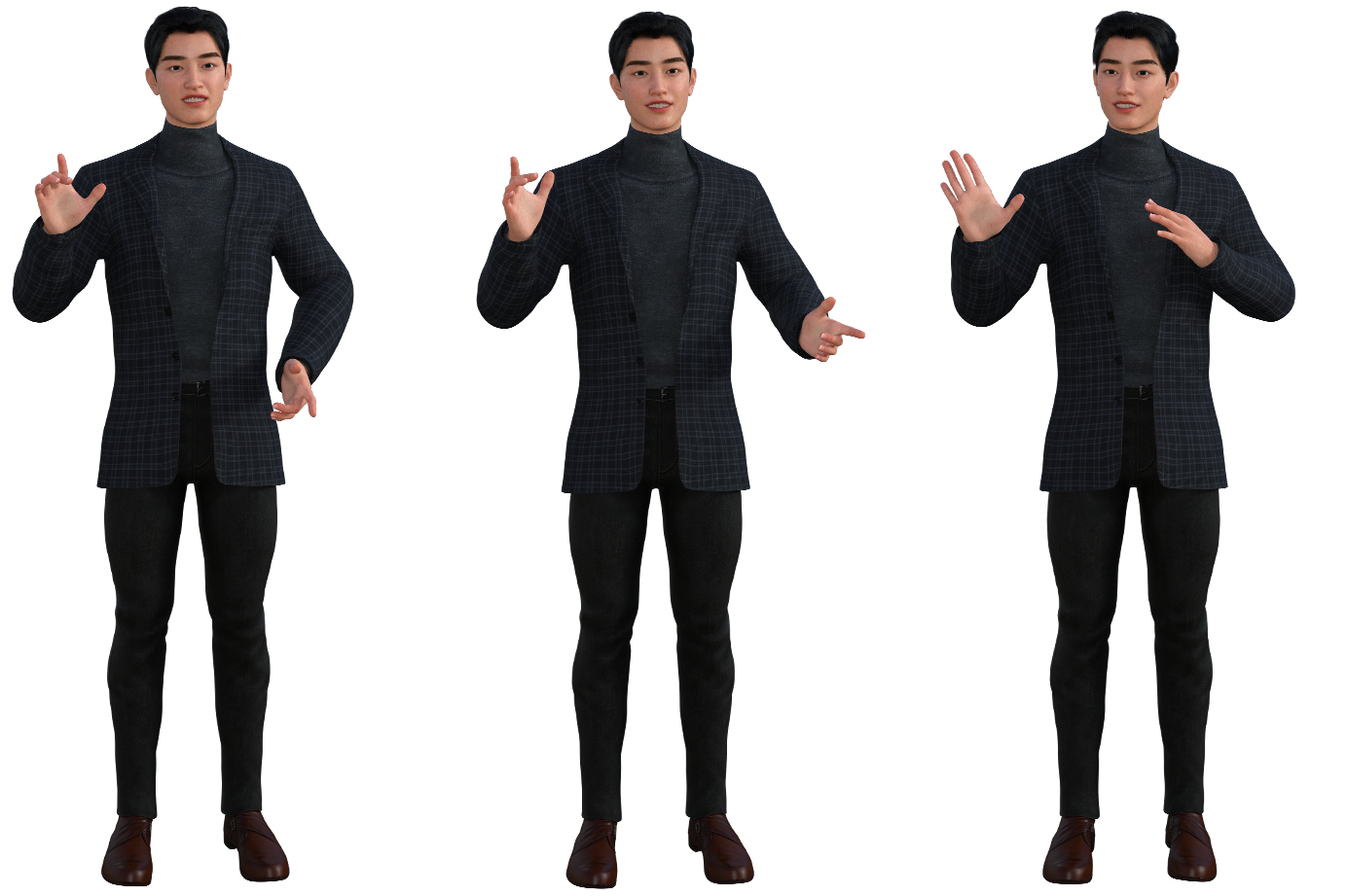}
    \caption{A male broadcaster is hosting a happy show (Left: GT; Center: DiM-Gestor; Right: PG-12blocks).}
    \label{fig:m_program_happy}
\end{figure}

Figures \ref{fig:f_program_calm} - \ref{fig:m_program_happy} illustrate the co-speech gesture effects generated by our proposed DiM-Gestor model, alongside a comparative analysis with the Ground Truth (GT) and the PG model. The results demonstrate that our model accurately captures the gesture styles corresponding to relevant emotions and contexts.

For instance, in a neutral-themed broadcast, the DiM-Gestor model generates gestures with the female announcer crossing her hands in front, reflecting a composed and professional demeanor. In contrast, the PG model depicts her with hands resting at her sides, missing the nuance of professional engagement (Figure \ref{fig:f_program_calm}). Similarly, when reciting content with a sad tone, the DiM-Gestor model positions the announcer's hands down bilaterally, conveying subdued emotions. In comparison, the PG model displays more excited gestures, which deviate from the intended emotional tone, as shown in Figure \ref{fig:f_recitation_sad}.

In another example, for male broadcasters, as illustrated in Figure \ref{fig:m_news_excited} and Figure \ref{fig:m_program_happy}, the GT data shows both hands bent forward, aligning with an expressive yet controlled delivery style. However, the PG model depicts hands pushed to the sides, failing to capture the gestural alignment with the speech context. These comparisons highlight the superior ability of the DiM-Gestor model to generate gestures that align with the emotional and contextual nuances of the speech, enhancing the authenticity and relatability of the digital human. 

\subsection{Subjective and Objective Evaluation}
In line with established practices in gesture generation research, we conducted a series of subjective and objective evaluations to assess the co-speech gestures generated by our proposed DiM-Gestor (DiM) model.

\begin{figure*}[!htbp]
  
    \centering
    \subfloat[Human-likeness]{\includegraphics[width=0.333\textwidth]{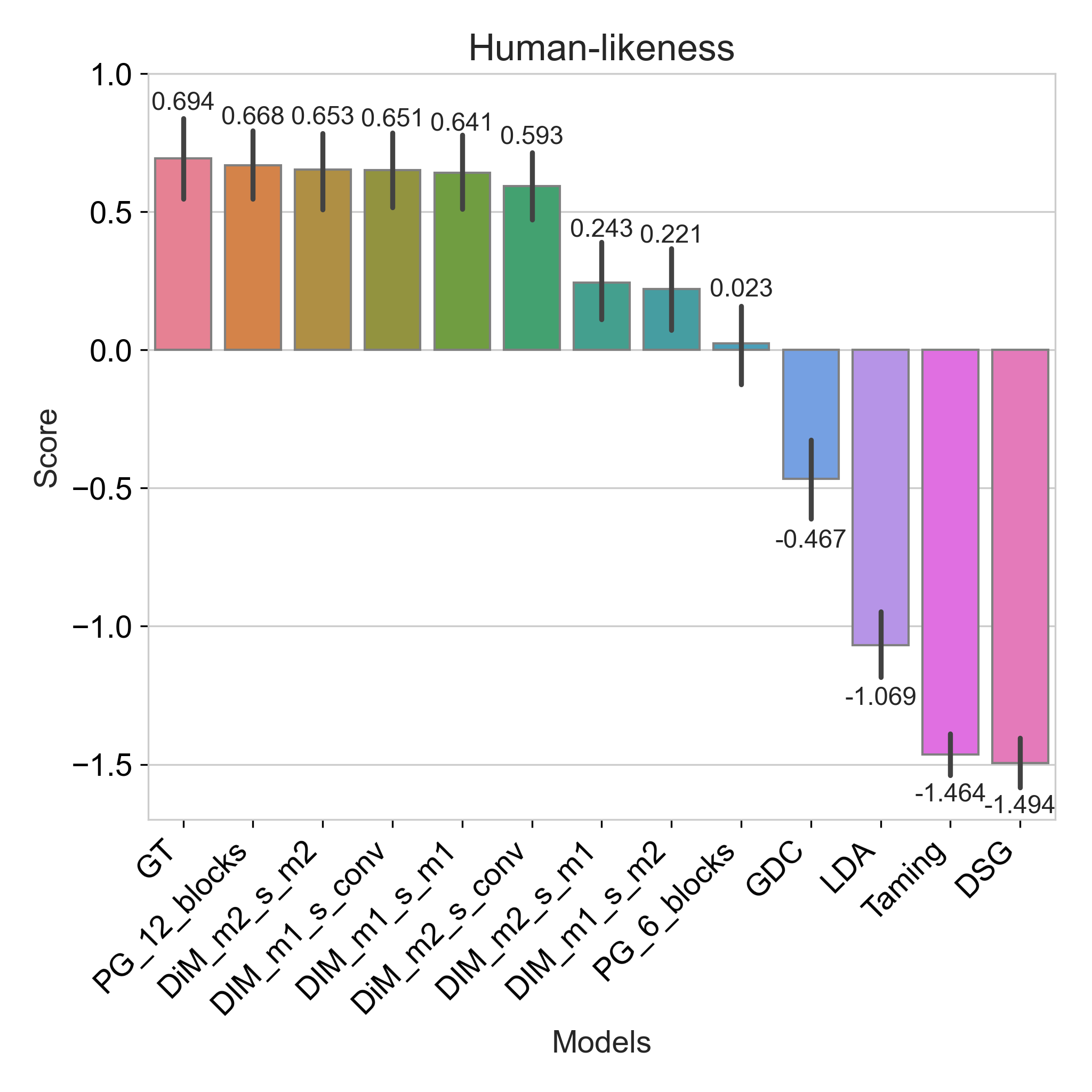}%
    \label{fig:HL} }
    % \hfil
    \subfloat[Appropriateness]{\includegraphics[width=0.333\textwidth]{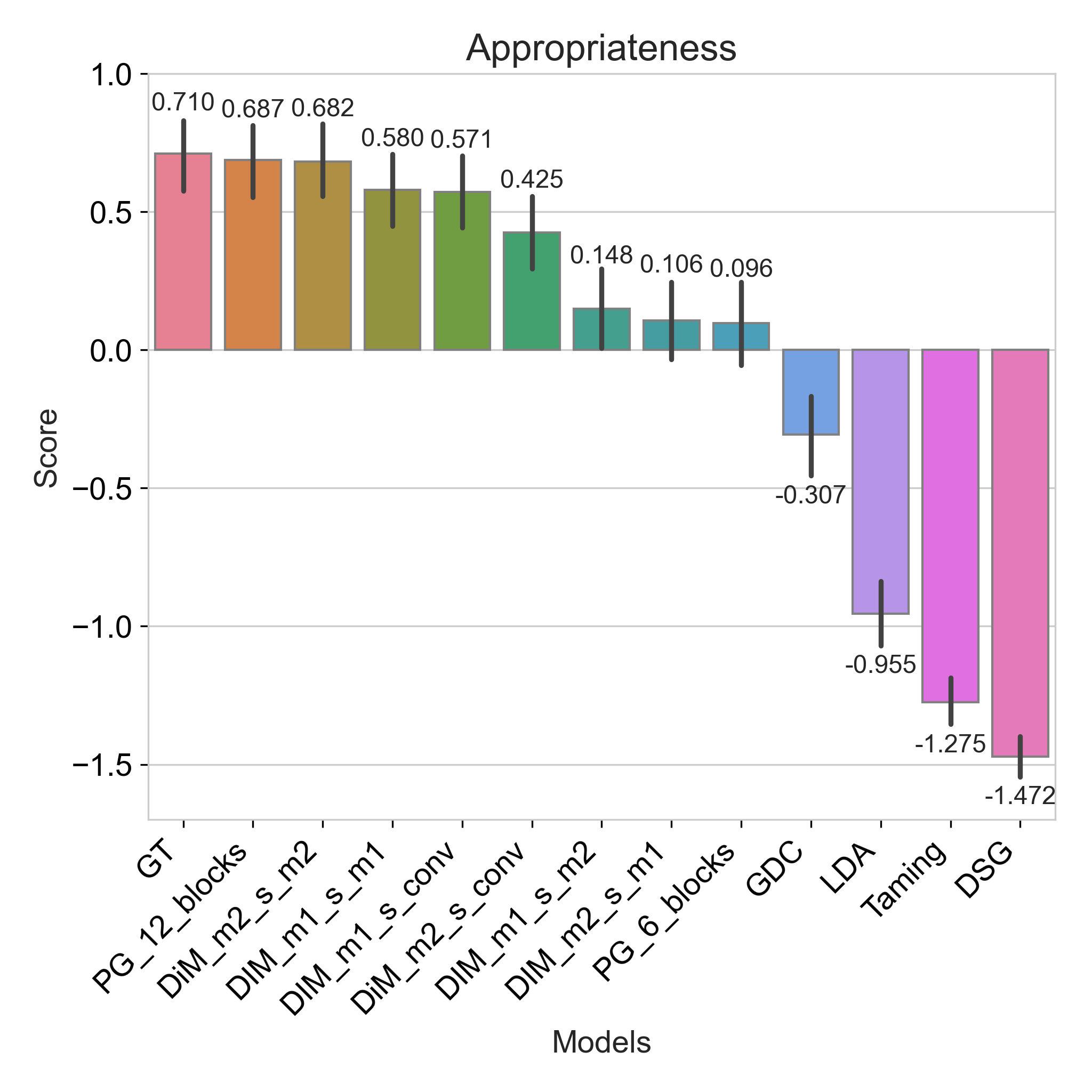}%
    \label{fig:APP} }
    % \hfil
    \subfloat[Style-Appropriateness]{\includegraphics[width=0.333\textwidth]{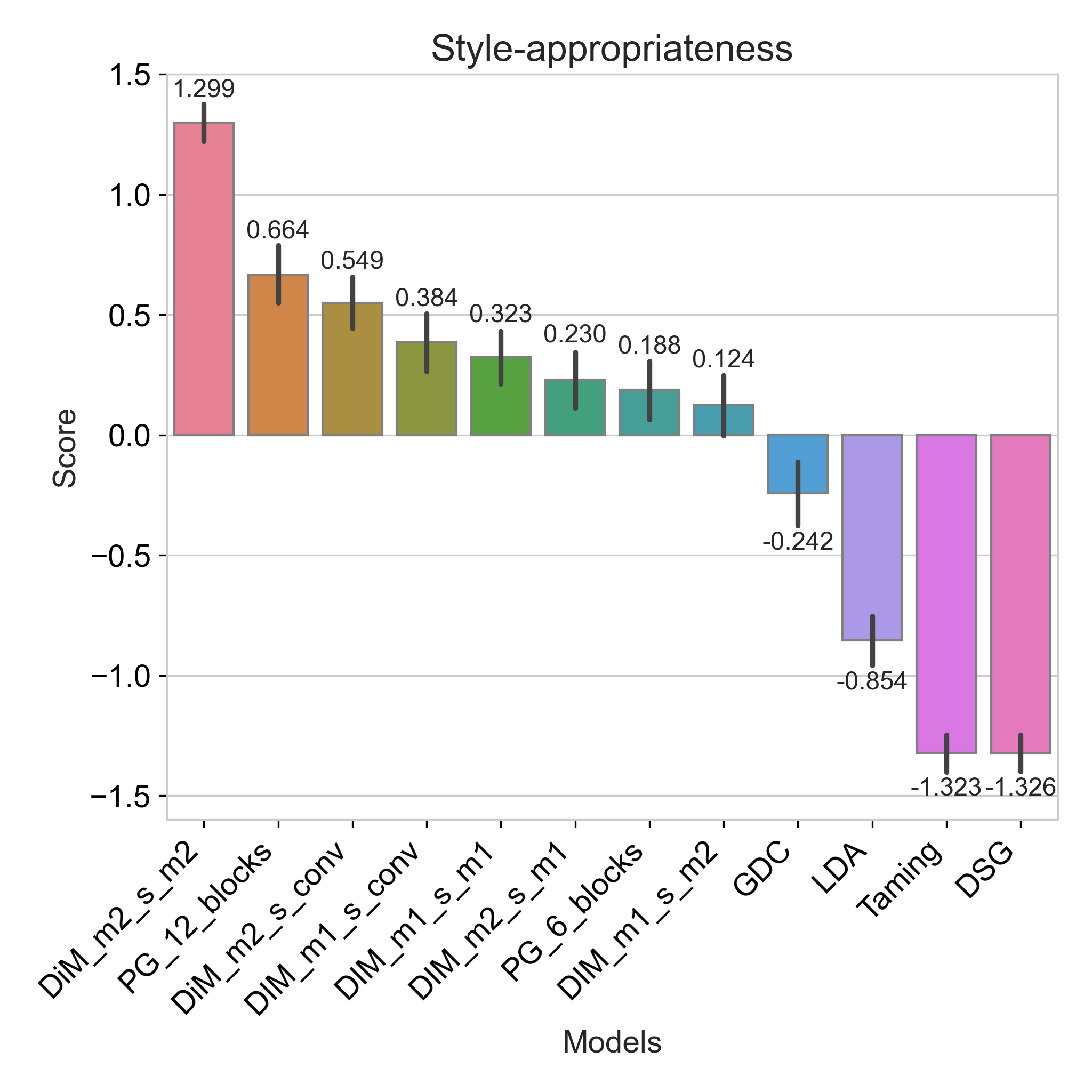}%
    \label{fig:StyleAPP} }
    \caption{The average metric rating for each approach in comparative experiments.}
\label{fig: histogram}
\end{figure*}

We primarily carried out experiments and made comparisons with models utilizing the transformers architecture, including LDA\cite{alexanderson2023listen}, DiffuseStyleGesture (DSG+)\cite{yang2023DiffuseStyleGestureaplus}, Taming\cite{zhu_taming_2023}, GDC\cite{ao2023GestureDiffuCLIP} and Persona-Gesture (PG) \cite{zhang_speech-driven_2024} as baseline models. The DSG++ are the best-performing in the 2023\cite{kucherenko_genea_2023} GENEA Challenge. We employ mamba-2 as a fuzzy feature extractor and adaLN mamba-2 architecture, denoted as DiM\_m2\_s\_m2, to abbreviate the model. The first 'm2' signifies the adoption of adaLN mamba-2, while the second 'm2'(after s) indicates the utilization of mamba-2 architecture for the fuzzy feature extractor. The PG model configured with 12 blocks incorporates a stack of 12 AdaLN transformer blocks, whereas the variant with 6 blocks consists of a stack with 6 AdaLN transformer blocks.

The baseline models were initially trained using English speech datasets, including Trinity\cite{ferstlylva2018Investigating}, ZEGGS\cite{ghorbani2023ZeroEGGS}, and BEAT\cite{liu2022BEAT}. Contrary to these settings, we employed our internally recorded Chinese dataset for training and inference. 

\subsubsection{Subjective Evaluation}
We utilize three distinct metrics for comprehensive subjective evaluations: human-likeness, appropriateness, and style-appropriateness. \textbf{Human-likeness} evaluates the naturalness and resemblance of the generated gestures to authentic human movements, independent of speech. \textbf{Appropriateness} assesses the temporal alignment of gestures with the speech's rhythm, intonation, and semantics to ensure a seamless and natural interaction. \textbf{Style-appropriateness} quantifies the similarity between the generated gestures and their original human counterparts, ensuring fidelity to the intended gesture style.

We executed a user study utilizing pairwise comparisons by the methodology outlined by \cite{wolfert_rate_2021}. During each session, participants were shown two 20-second video clips. These clips, generated by different models, including the Ground Truth (GT), were presented for direct comparative analysis. Participants were instructed to select the preferred clip based on predefined evaluation criteria. Preferences were quantified on a scale ranging from 0 to 2, with the non-selected clip in each pair receiving a corresponding inverse score. A score of zero was used to indicate a neutral preference. More details are described in \cite{wolfert_rate_2021, zhang_speech-driven_2024}.

Given the extensive array of styles within the datasets, an individual evaluation of each style was considered unfeasible. A random selection methodology was employed to address this. Each participant was assigned a subset of five styles for assessment. Critically, none of the audio clips selected for evaluation were used in the training or validation sets. This strategy ensured a broad yet manageable coverage of the dataset's diversity in a controlled and unbiased manner.

\begin{figure*}[!htbp]
    
    \centering
    \subfloat[Human-likeness]{\includegraphics[width=0.333\textwidth]{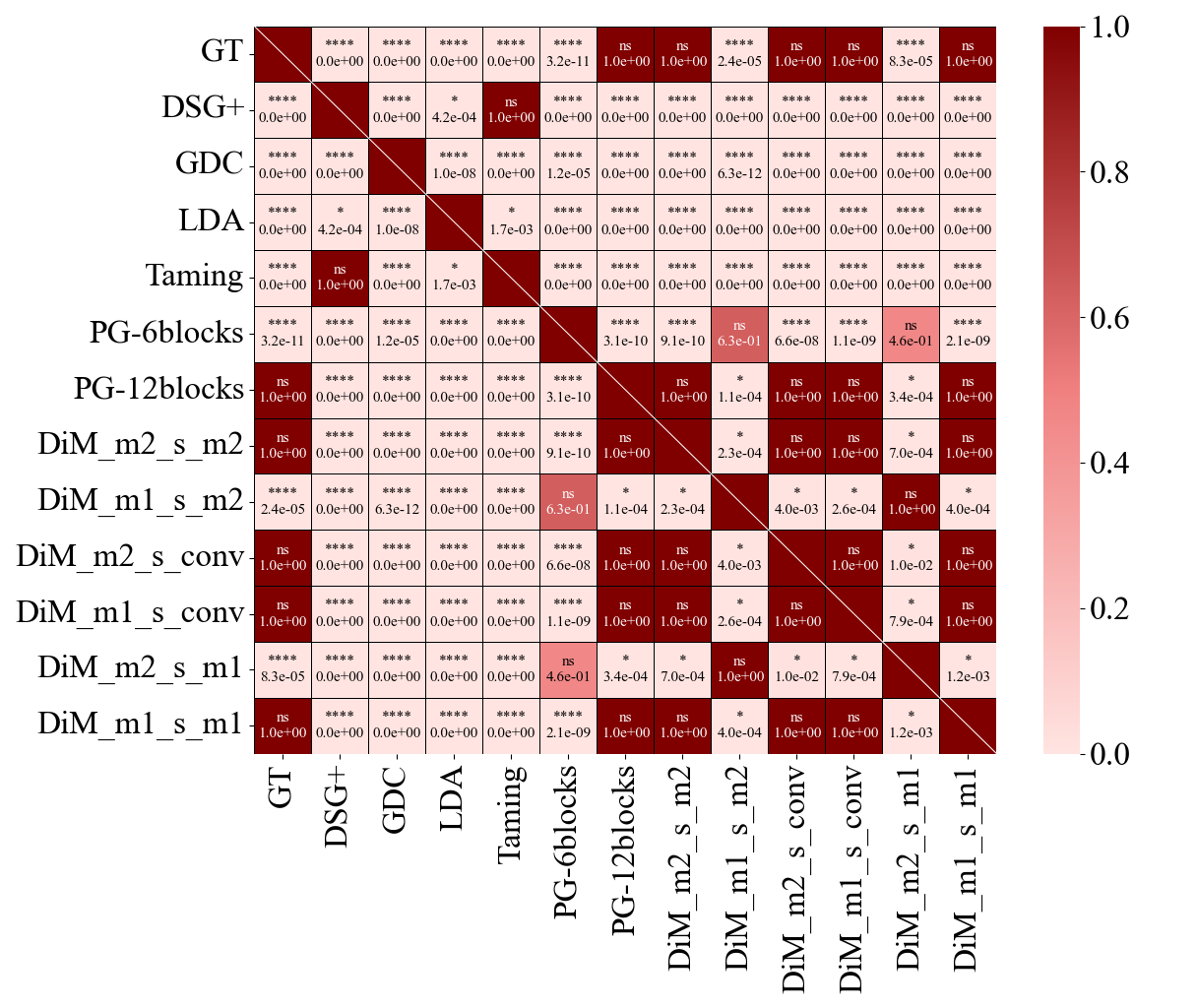}%
    \label{fig:HLheatmap} }
    % \hfil
    \subfloat[Appropriateness]{\includegraphics[width=0.333\textwidth]{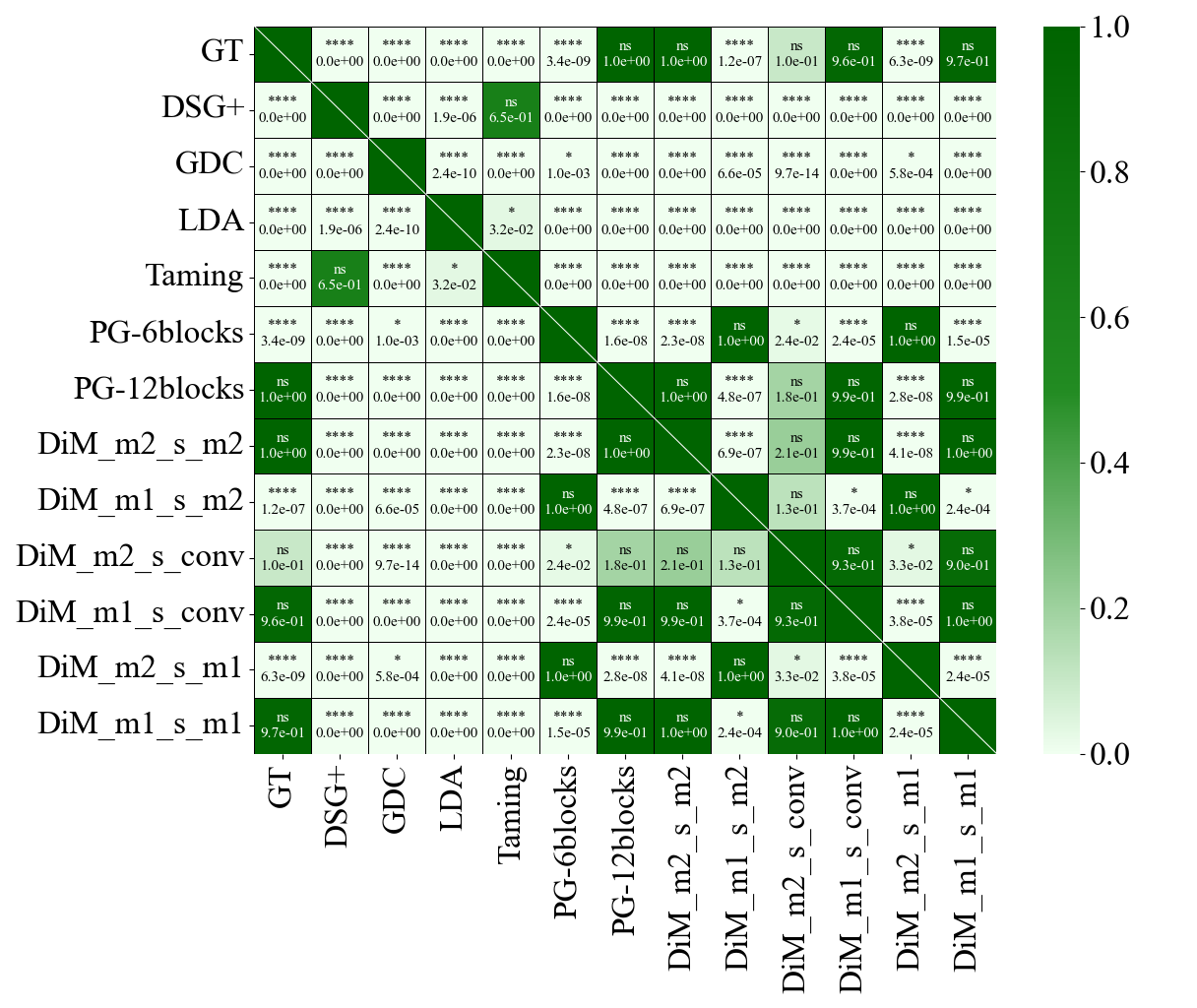}%
    \label{fig:APPheatmap} }
    % \hfil
    \subfloat[Style-Appropriateness]{\includegraphics[width=0.333\textwidth]{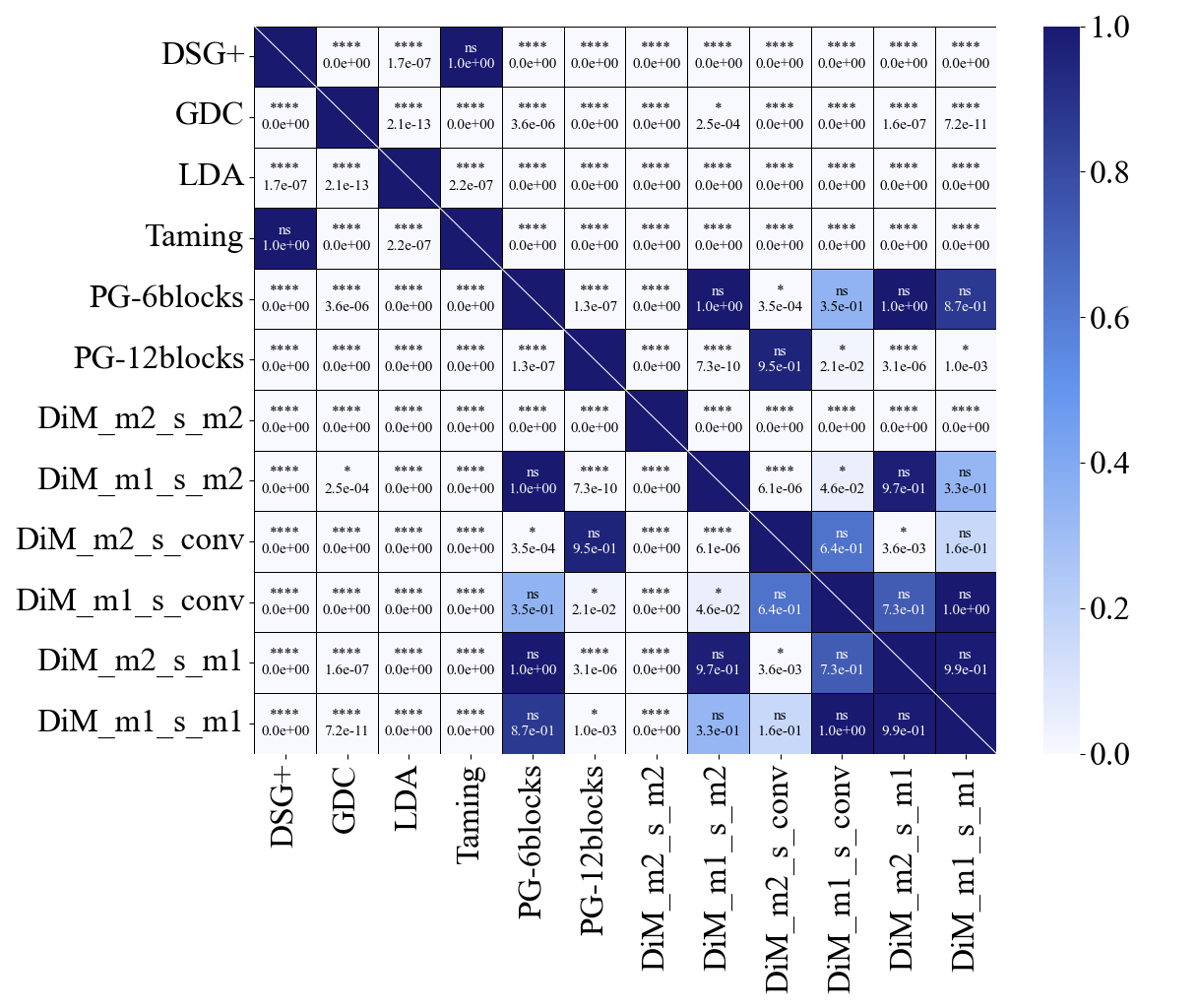}%
    \label{fig:StyleAPPheatmap} }
    \caption{Heatmaps of the mean ratings of user studies significant differences across all methods for each metric. Asterisks indicated significant effects (*: $p < 0.05$, ****: $p < 1.00e-04$ , ns: no significant difference). We use distinct colors to represent each metric: red for human likeness, green for Appropriateness, and blue for Style Appropriateness, with lighter shades indicating greater significance differences.}
\label{fig: heatmap}
\end{figure*}

We invited 30 native Chinese volunteers—14 males and 16 females aged between 18 and 35 for user study. 

One-way ANOVA and post hoc Tukey, multiple comparison tests, were conducted to determine if there were statistically significant differences among the models' scores across the three evaluation aspects. The results are presented in Figure \ref{fig: histogram} and Table \ref{tab:overview}, offering detailed insights into the performance variances observed among models regarding human-likeness, appropriateness, and style-appropriateness. 

\begin{table}[htbp]
\caption{The subject mean perceptual rating score. Bold and red fonts were utilized to emphasize the best results for each metric among the different methods, except for the GT.}
\label{tab:overview}
\centering
\renewcommand{\arraystretch}{1.2}
\resizebox{\linewidth}{!}{
\begin{tabular}{ccccc}
\hline
\multicolumn{5}{c}{\bf{Subject Evaluation Metrics}}                                                                                                                                                                                                                                      \\ \hline
\bf{Model}                                                               & \begin{tabular}[c]{@{}c@{}}\bf{With}\\ \bf{Fingers}\end{tabular} & \begin{tabular}[c]{@{}c@{}}\bf{Human↑}\\ \bf{likeness}\end{tabular} & \bf{Appropriateness↑}     & \begin{tabular}[c]{@{}c@{}}\bf{Style↑}\\ \bf{appropriateness}\end{tabular} \\ \hline
GT                                                                  & Y                                                      & 0.694±1.430                                               & 0.710±1.268          & /                                                                \\
DSG+                                                                & Y                                                      & -1.494±0.794                                              & -1.472±0.712         & -1.326±0.744                                                     \\
GDC                                                                 &  \textit{N}                     & -0.467±1.399                                              & -0.307±1.394         & -0.242±1.306                                                     \\
LDA                                                                 & \textit{N}                     & \multicolumn{1}{l}{-1.069±1.099}                          & -0.955±1.076         & -0.854±1.027                                                     \\
Taming                                                              & Y                                                      & -1.464±0.692                                              & -1.276±0.796         & -1.323±0.769                                                     \\ \hline
PG-6blocks                                                          & Y                                                      & 0.023±1.390                                               & 0.096±1.376          & 0.188±1.188                                                      \\
PG-12blocks                                                         & Y                                                      & \multicolumn{1}{l}{{\color[HTML]{FE0000}\textit{\textbf{0.668±1.192}}}}                  & {\color[HTML]{FE0000}\textit{\textbf{0.687±1.247}}} & 0.664±1.216                                                      \\ \hline
\begin{tabular}[c]{@{}c@{}}DiM\_m2\_s\_m2(Ours)\\ \end{tabular} & Y                                                      & \multicolumn{1}{l}{0.653±1.323}                           & 0.682±1.326          & {\color[HTML]{FE0000}\textit{\textbf{1.30±0.770}}}                                              \\
DiM\_m1\_s\_m2                                                      & Y                                                      & \multicolumn{1}{l}{0.221±1.355}                           & 0.148±1.327          & 0.124±1.139                                                      \\
DiM\_m2\_s\_conv                                                    & Y                                                      & 0.593±1.222                                               & 0.425±1.319          & 0.549±1.111                                                      \\
DiM\_m1\_s\_conv                                                    & Y                                                      & 0.651±1.279                                               & 0.571±1.297          & 0.384±1.186                                                      \\
DiM\_m2\_s\_m1                                                      & Y                                                      & \multicolumn{1}{l}{0.243±1.348}                           & 0.106±1.420          & 0.230±1.221                                                      \\
DiM\_m1\_s\_m1                                                      & Y                                                      & \multicolumn{1}{l}{0.641±1.236}                           & 0.580±1.280          & 0.323±1.136                                                      \\ \hline
\end{tabular}
}
\end{table}

Regarding the Human Likeness metric, our proposed model, \textit{DiM\_m2\_s\_m2}, achieves a score of \(0.653 \pm 1.323\). This result closely approximates the top-performing PG-12blocks model (\(0.668 \pm 1.192\)) and the Ground Truth (GT) benchmark (\(0.694 \pm 1.430\)). Statistical analysis revealed no significant differences among these three models, as Figure \ref{fig: heatmap} illustrates. This indicates that the \textit{DiM\_m2\_s\_m2} model performs comparably to the human baseline regarding perceived naturalness. In contrast, alternative methods such as DSG+ (\(-1.494 \pm 0.794\)), GDC (\(-0.467 \pm 1.399\)), and LDA (\(-1.069 \pm 1.099\)) exhibit significantly lower scores on the Human Likeness metric.

In terms of the Appropriateness metric, \textit{DiM\_m2\_s\_m2} attains a score of \(0.682 \pm 1.326\), demonstrating high competitiveness and being nearly on par with PG-12blocks, the top-performing synthetic model with a score of \(0.687 \pm 1.247\). The Ground Truth (GT) establishes the benchmark at a score of \(0.710 \pm 1.268\). Additionally, no significant difference exists among these three models. The models' significant differences are visually depicted in Figure \ref{fig: heatmap}. Models such as DSG+, Taming, and GDC, which record scores within the negative range, evidently have difficulty synchronizing gestures with the speech context.

The model \textit{DiM\_m2\_s\_m2} exhibits significant superiority in the Style Appropriateness metric, achieving the highest score of \(1.30 \pm 0.770\) among all evaluated models. This metric underscores our method's capability to generate gestures stylistically congruent with the various Chinese speech styles. In contrast, the PG-12blocks model attains a lower score of \(0.664 \pm 1.216\), while the other methods, including DSG+, GDC, and LDA, exhibit negative scores in this category. These findings emphasize the distinct advantage of DiM\_m2\_s\_m2 in producing gestures that align closely with the intended stylistic attributes of the spoken language.

In conclusion, \textit{DiM\_m2\_s\_m2} outperforms alternative models in Style Appropriateness and achieves highly competitive results in Human-Likeness and Appropriateness. These findings suggest that \textit{DiM\_m2\_s\_m2} effectively generates perceptually realistic gestures and is well-aligned with speech-driven gesture synthesis's contextual and stylistic requirements. This highlights the strength of our Mamba-2 approach in addressing the multi-dimensional challenges in this domain, setting a new standard for synthetic gesture quality compared to traditional transformer methods.

\subsubsection{Objective Evaluation}\label{sec:objective eval}

We employ three objective evaluation metrics to assess the quality and synchronization of generated gestures: \textit{Fréchet Gesture Distance (FGD)} in both feature and raw data spaces\cite{yoonyoungwoo2020Speecha}, and \textit{BeatAlign}\cite{li2021ai}. Inspired by the Fréchet Inception Distance (FID)\cite{heusel2017gans}, \textbf{FGD} evaluates the quality of generated gestures and has shown moderate correlation with human-likeness ratings, surpassing other objective metrics\cite{kucherenko2023Evaluating}. \textbf{BeatAlign}, on the other hand, assesses gesture-audio synchrony by calculating the Chamfer Distance between audio beats and gesture beats, thus providing insights into the temporal alignment of gestures with speech rhythms.

\begin{table}[ht]
\caption{The objective score. Bold and Red fonts were utilized to emphasize the best results for each metric among the different methods, except for the GT.}
\label{tab:objective}
\centering
\renewcommand{\arraystretch}{1.2}
\resizebox{\linewidth}{!}{
\begin{tabular}{cccc}
\hline
\multicolumn{4}{c}{\bf{Objective Evaluation Metrics}}                                                                                                                                                                                 \\ \hline
\bf{Model}                                                               & \begin{tabular}[c]{@{}c@{}}\bf{FGD↓}\\ \bf{on feature space}\end{tabular} & \begin{tabular}[c]{@{}c@{}}\bf{FGD↓}\\ \bf{on raw data space}\end{tabular} & \bf{BeatAlign↑}           \\ \hline
GT                                                                  & /                                                           & /                                                            & /          \\
DSG+                                                                & 42329.558                                                       & 12206710.224                                                     & 0.423          \\
GDC                                                                 & 148.937                                                         & 2536.248                                                         & 0.657         \\
LDA                                                                 & 34678.668                                                       & 12305524.751                                                     & 0.428          \\
Taming                                                              & 5699.738                                                        & 394606.850                                                       & 0.667          \\ \hline
PG-6blocks                                                          & 108.913                                                         & 1952.618                                                         & 0.667          \\
PG-12blocks                                                         & 100.899                                                         & 2128.274                                                         & 0.669          \\ \hline
\begin{tabular}[c]{@{}c@{}}{DiM\_m2\_s\_m2 (Ours)}\\ \end{tabular} & {\color[HTML]{FE0000}\textit{\textbf{17.716}}}                                                 & {\color[HTML]{FE0000}\textit{\textbf{424.287}}}                                                 & 0.670          \\
DiM\_m1\_s\_conv                                                    & 87.025                                                          & 1293.644                                                         & 0.672          \\
DiM\_m2\_s\_conv                                                    & 44.282                                                          & 862.673                                                          & {\color[HTML]{FE0000}\textit{\textbf{0.678}}} \\
DiM\_m1\_s\_m1                                                      & 130.119                                                         & 1789.393                                                         & 0.669          \\
DiM\_m2\_s\_m1                                                      & 120.053                                                         & 1911.956                                                         & 0.669          \\
DiM\_m1\_s\_m2                                                      & 184.300                                                         & 2174.291                                                         & 0.674          \\
 \hline
\end{tabular}
}
\end{table}

Table \ref{tab:objective} provides a comparison of the objective evaluation metrics for several models in speech-driven gesture synthesis, including our proposed method, \textit{DiM\_m2\_s\_m2}, alongside GT, DSG+, GDC, LDA, Taming, PG-6blocks, and PG-12blocks. The metrics evaluated are the Fréchet Gesture Distance (FGD) in both feature space and raw data space, as well as the BeatAlign score. Lower FGD values and higher BeatAlign scores indicate better performance.

Our proposed model, \textit{DiM\_m2\_s\_m2}, achieves the lowest FGD on the feature space among all synthetic models with a score of 17.716, significantly outperforming the other methods. By comparison, PG-12blocks and PG-6blocks, which also performed well on other perceptual metrics, scored 100.899 and 108.913, respectively. High FGD values in DSG+ (42329.558) and LDA (34678.668) indicate poor alignment with the target gesture distribution, highlighting our approach's substantial improvements in naturalness and similarity to natural gestures.

Regarding the BeatAlign metric, \textit{DiM\_m2\_s\_m2} scores competitively with a BeatAlign value of 0.670. While this is close to the highest scores achieved by PG-12blocks (0.669) and DiM\_m2\_s\_conv (0.678), it demonstrates the overall balance of our model across both spatial and temporal metrics. Models such as DSG+ (0.423) and LDA (0.428) scored significantly lower, indicating suboptimal temporal synchronization.

In summary, \textit{DiM\_m2\_s\_m2} achieves superior performance across multiple objective metrics, with the lowest FGD values in both feature and raw data spaces, indicating a close resemblance to natural gestures. Although its BeatAlign score is slightly lower than the top-performing PG-12blocks, the overall results validate \textit{DiM\_m2\_s\_m2} as a highly effective approach for generating temporally and spatially consistent gestures in the domain of speech-driven gesture synthesis.

\subsection{Ablation Studies}

This section details an ablation study to assess the individual contributions of two key components in our proposed model: the \textit{fuzzy feature extractor} and the \textit{AdaLN architectures} equipped with different versions of the Mamba framework, as shown in Tables \ref{tab:overview}, \ref{tab:objective}, and Figure \ref{fig: histogram}. 
 
\subsubsection{Effect of Fuzzy Feature Extractor}
Table \ref{tab:overview} and Figure \ref{fig: histogram} present the performance differences when utilizing \textit{mamba-1}, \textit{mamba-2}, and \textit{convolution 1D} \cite{zhang_speech-driven_2024} as the fuzzy feature extractor across various model configurations. In all instances, the AdaLN module is integrated with the Mamba-2 architecture. 

In the Human-likeness metric, both \textit{DiM\_m2\_s\_m2} (0.653 ± 1.323) and DiM\_m2\_s\_conv (0.593 ± 1.222) achieve high scores, with no statistically significant difference ($p = 1.0 > 0.05$) observed between the two. These findings suggest that Mamba-2 and 1D Convolution represent competitive alternatives for capturing human-like gestures. However, the model DiM\_m2\_s\_m1, which employs \textit{Mamba-1}, achieves a significantly lower score of \(0.243 \pm 1.348\).

Similarly, in the Appropriateness metric, there is no significant difference ($p = 2.1e-1 > 0.05$) between \textit{DiM\_m2\_s\_m2} (0.682 ± 1.326) and DiM\_m2\_s\_conv (0.425 ± 1.319). Both models align well with the contextual requirements of speech-driven gestures. However, DiM\_m2\_s\_m1 achieves a much lower score (0.106 ± 1.420).

In Style Appropriateness, \textit{DiM\_m2\_s\_m2} outperforms the alternatives, achieving the highest score of 1.30 ± 0.770. This result underscores the superiority of \textit{Mamba-2} in capturing stylistic nuances and generating gestures that are contextually relevant and visually coherent with the speech content. In comparison, DiM\_m2\_s\_conv scores moderately at 0.549 ± 1.111, while DiM\_m2\_s\_m1 achieves a significantly lower score of 0.230 ± 1.221.

The objective evaluation results in Table \ref{tab:objective} highlight the benefits of using the \textit{mamba-2} fuzzy feature extractor in achieving superior alignment and detail in gesture synthesis. Specifically, \textit{DiM\_m2\_s\_m2} records the lowest Fréchet Gesture Distance (FGD) in feature space at 17.716, outperforming alternative configurations like DiM\_m2\_s\_conv (44.282) and DiM\_m2\_s\_m1 (120.053). 

For the BeatAlign metric, which measures temporal synchronization with speech, DiM\_m2\_s\_conv achieves the highest score of 0.678, slightly outperforming \textit{DiM\_m2\_s\_m2} (0.670) and DiM\_m2\_s\_m1 (0.669). While \textit{mamba-2} in DiM\_m2\_s\_m2 remains competitive.

The ablation study of fuzzy feature extractor demonstrates that our proposed model, DiM\_m2\_s\_m2, utilizing \textit{Mamba-2} for the fuzzy feature extractor, consistently outperforms the \textit{Mamba-1} configurations, particularly in Style Appropriateness. Although no significant difference was observed between DiM\_m2\_s\_m2 and DiM\_m2\_s\_conv in Human Likeness and Appropriateness, DiM\_m2\_s\_m2 demonstrates a clear advantage in Style Appropriateness. These findings validate the effectiveness of \textit{Mamba-2} in achieving high-quality gesture synthesis that balances naturalness, contextual alignment, and stylistic coherence.

\subsubsection{Effect of adaLN mamba Architecture}
This experiment investigates the impact of employing different adaLN Mamba architectures (adaLN Mamba-1 and adaLN Mamba-2) on the generation effect, as shown in Table \ref{tab:overview} and Table \ref{tab:objective}. In the Human Likeness metric, \textit{DiM\_m2\_s\_m2} achieves a high score of 0.653 ± 1.323, comparable to DiM\_m1\_s\_conv (0.651 ± 1.279) and DiM\_m1\_s\_m1 (0.641 ± 1.236). Statistical analysis reveals no significant difference between these models in this metric (Figure \ref{fig: histogram}). However, DiM\_m2\_s\_m1, which uses the \textit{Mamba-1} architecture, scores considerably lower at 0.243 ± 1.348. This suggests that while both architectures can produce human-like gestures, the \textit{Mamba-2} architecture in DiM\_m2\_s\_m2 exhibits slight improvements in capturing nuanced human motion.

Style Appropriateness highlights a more pronounced distinction between the models. Our proposed model, \textit{DiM\_m2\_s\_m2}, achieves the highest score of 1.30 ± 0.770, indicating superior stylistic coherence and visual appeal in gesture synthesis. Both DiM\_m1\_s\_conv (0.384 ± 1.186) and DiM\_m1\_s\_m1 (0.323 ± 1.136) perform moderately, while DiM\_m2\_s\_m1 again scores poorly (0.230 ± 1.221).

Quantitatively, Table \ref{tab:objective} highlights that models with \textit{adaLN mamba-2} architectures, like \textit{DiM\_m2\_s\_m2}, consistently achieve lower FGD scores, underscoring their alignment with natural gestures. For instance, DiM\_m2\_s\_conv records an FGD of 44.282 in feature space, while DiM\_m1\_s\_m1, using \textit{adaLN mamba-1}, records a significantly higher FGD of 130.119. This suggests that \textit{adaLN mamba-2} enhances alignment with the target gesture distribution.

The ablation study confirms that using \textit{mamba-2} for both the fuzzy feature extractor and adaLN mamba architecture provides optimal results. Our proposed model, \textit{DiM\_m2\_s\_m2}, outperforms all other configurations in perceptual and quantitative metrics, highlighting the combined benefits of the \textit{mamba-2} configurations in generating realistic, contextually appropriate, and stylistically aligned gestures in speech-driven gesture synthesis. 

Interestingly, experiments that mixed Mamba-1 and Mamba-2 architectures resulted in a noticeable decrease in performance. Conversely, utilizing a consistent Mamba architecture throughout substantially enhanced the scores. The best results were achieved using Mamba-2, underscoring its superior effectiveness in this application context.
\section{Paramter Counts and Inference Speed}
This section discusses the parameter counts and inference speed among DiM-Gestor with different Mamba versions and the PG, which utilizing AdaLN transformer architecture.
\subsection{Parameter Counts}
\textit{DiM\_m2\_s\_m2} demonstrates a relatively low parameter count of 535M, making it considerably more compact than configurations such as DiM\_m1\_s\_conv (836M) and DiM\_m2\_s\_conv (826M). This compactness is advantageous, as it indicates that \textit{DiM\_m2\_s\_m2} can achieve high-quality gesture synthesis with reduced computational overhead compared to larger models. While the AdaLN-based transformer model, PG-12blocks, can generate high-quality gestures, its significantly larger parameter count of 1.2B may impact memory requirements, posing challenges for deployment in resource-constrained environments. 

These observations highlight the efficiency and practicality of \textit{DiM\_m2\_s\_m2} as a robust yet computationally lightweight solution for speech-driven gesture synthesis.

\subsection{Inference Speed}
\begin{figure}[!htbp]
    \centering
    \includegraphics[width=\linewidth]{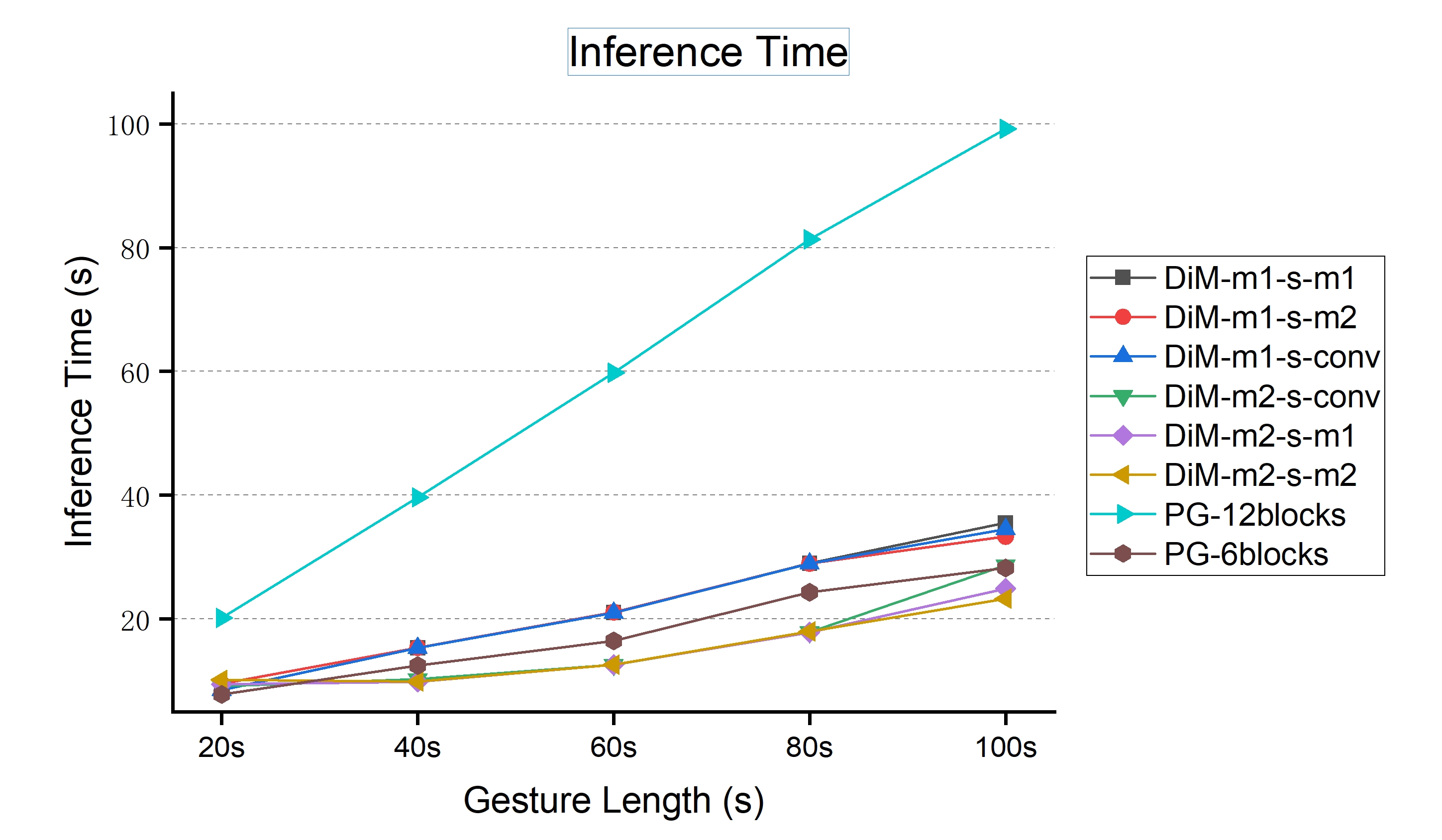}
    \caption{The inference time of various gesture lengths across different architectures.}
    \label{fig: InferenceTime}
\end{figure}

We aggregated the 20s audio segments into durations of 40s, 60s, 80s, and 100s to evaluate the computational efficiency of different models in processing gesture sequences with varying lengths. Table \ref{tab:ablation} and Figure \ref{fig: InferenceTime} also detail each model configuration's inference times across different gesture sequence lengths. \textit{DiM\_m2\_s\_m2} shows a balanced performance, with inference times of 10.16s for 20s sequences and scaling to 23.27s for 100s sequences. This model maintains competitive inference speed even for longer sequences, outperforming other configurations with similar parameter sizes. For example, PG-12blocks requires 20.153 seconds of inference time for generating a 20-second gesture sequence, which scales to 40 seconds for a 40-second sequence and 100 seconds for a 100-second sequence. This highlights the need for the computational resources of the PG-12blocks model. 

\begin{table}[!htbp]
\caption{In the ablation study focusing on the Mamba-based style fuzzy feature extractor and the AdaLN Mamba2, we meticulously assessed the parameter count and inference time to quantify these components' efficiency and performance impact.}
\resizebox{\linewidth}{!}{
\label{tab:ablation}
\renewcommand{\arraystretch}{1.2}
\begin{tabular}{ccccccc}
\hline
\multicolumn{7}{c}{\bf{Parameter Counts and Inference Time}}                                                                                           \\ \hline
\multirow{2}{*}{\bf{Models}} & \multirow{2}{*}{\begin{tabular}[c]{@{}c@{}}\bf{Param.↓}\\ \bf{Counts}\end{tabular}} & \multicolumn{5}{c}{\bf{Gesture Lengths (Second)}} \\ \cline{3-7} 
                        &                                                                          & \bf{20s↓}     & \bf{40s↓}     & \bf{60s↓}    & \bf{80s↓}    & \bf{100s↓}   \\ \hline
PG-6blocks              & 812M                                                                     & \color[HTML]{FE0000}\textit{\textbf{7.87}}    & 12.46   & 16.44  & 24.31  & 28.22  \\
PG-12blocks             & 1.2B                                                                     & 20.153& 39.637& 59.785& 81.344& 99.214\\ \hline
DiM\_m2\_s\_m2(Ours)             & \color[HTML]{FE0000}\textit{\textbf{535M}}                                                                     & 10.16   & 9.86    & 12.60  & 17.96  & \color[HTML]{FE0000}\textit{\textbf{23.27}}  \\
DiM\_m1\_s\_m2             & 545M                                                                     & 9.51    & 15.34   & 21.08  & 28.95  & 33.32  \\
DiM\_m2\_s\_conv           & 826M                                                                     & 9.13    & 10.26   & \color[HTML]{FE0000}\textit{\textbf{12.58}}  & 17.84  & 28.67  \\
DiM\_m1\_s\_conv           & 836M                                                                     & 8.49    & 15.30   & 20.98  & 28.93  & 34.50  \\
DiM\_m2\_s\_m1             & 536M                                                                     & 9.43    & \color[HTML]{FE0000}\textit{\textbf{9.83}}    & 12.59  & \color[HTML]{FE0000}\textit{\textbf{17.81}}  & 24.89  \\
DiM\_m1\_s\_m1             & 546M                                                                     & 8.51    & 15.31   & 21.00  & 28.99  & 35.52  \\ \hline
\end{tabular}
}
\end{table}

Compared to the AdaLN Transformer architecture that requires 12 blocks, the DiM model employs only 6 AdaLN Mamba-2 blocks to achieve comparable performance in generating co-speech gestures. This configuration results in a reduction of memory usage by approximately 2.4 times and an increase in inference speeds by 2 to 4 times.

The results of this ablation study underscore the efficiency of our proposed model, \textit{DiM\_m2\_s\_m2}, which achieves a favorable balance between parameter count and inference speed. With a lower parameter count of 535M and competitive inference times, \textit{DiM\_m2\_s\_m2} is well-suited for applications requiring both performance and efficiency in speech-driven gesture synthesis. These findings highlight DiM\_m2\_s\_m2 as a compact and efficient model, outperforming other tested architectures, particularly in scenarios involving long-duration gesture synthesis tasks. 

In summary, our experimental findings confirm that DiM-Gestor offers competitive advantages over traditional Transformer-based models, achieving state-of-the-art results in style appropriateness and excelling in metrics of human-likeness and appropriateness. These outcomes validate the model's capability to produce visually convincing gestures and contextually synchronized with speech. Further, our ablation study highlights the Mamba-2 architecture's superior performance and efficiency, which is particularly beneficial for processing long sequences. Compared with PG with 12 12-block adaLN transformer, implementing this architecture significantly reduces memory usage—approximately 2.4 times—and increases inference speeds by 2 to 4 times, underscoring its potential for scalable and real-time applications.

\section{DISSCUSTION and CONCLUSION}\label{sec:CONCLUSION}
This study introduces DiM-Gestor, a novel architecture for co-speech gesture synthesis, employing a Mamba-2 fuzzy feature extractor and AdaLN Mamba-2 within a diffusion framework. This model is designed to generate highly personalized 3D full-body gestures solely from raw speech audio, representing a significant advancement in gesture synthesis technologies for virtual human applications.

DiM-Gestor’s fuzzy feature extractor leverages Mamba-2 to capture implicit, speaker-specific features from audio. These synthesizing gestures resonate with the style and rhythm of the speaker’s voice without relying on predefined style labels. This approach improves generalization and usability, making it applicable across diverse scenarios without requiring extensive label-specific training data. The integration of AdaLN Mamba-2 further enhances the model’s efficiency and flexibility, surpassing traditional transformer-based methods regarding memory efficiency and inference speed. AdaLN Mamba-2 reduces computational overhead by maintaining linear complexity, making real-time applications feasible while maintaining gesture quality comparable to the existing Transformer-based Persona-Gestor.

In addition to model advancements, we contribute a large-scale, high-quality Chinese Co-Speech Gesture Dataset (CCG dataset), recorded by professional broadcasters, encompassing various styles and scenarios. This dataset enriches the resources available for research and development in this field, particularly for applications involving formal and structured speech in the Chinese language.

For future developments, specifically by integrating the accelerated diffusion model sCM \cite{lu_simplifying_2024}. This adaptation could amplify inference speeds by at least 50 times, catering to interactive applications' stringent real-time performance requirements. This enhancement would refine the user experience and broaden the model's applicability across various dynamic and user-centric platforms.

% \section{ACKNOWLEDGMENTS}

\bibliography{mybibliography}

\bibliographystyle{IEEEtran}

\end{document}